\documentclass[trackchanges,twocolumn]{aastex701}
\usepackage{siunitx}

\begin{document}

\title{Assessing Ultra-Cool Dwarf Contamination in Photometrically Selected High-Redshift Galaxy Samples}

\author[0009-0006-0994-1706]{Onnalin Innala}
\affiliation{National Astronomical Research Institute of Thailand (NARIT), Mae Rim, Chiang Mai, 50180, Thailand}
\email{onnalin@narit.or.th}

\author[0000-0003-4570-3159]{Nicha Leethochawalit}
\affiliation{National Astronomical Research Institute of Thailand (NARIT), Mae Rim, Chiang Mai, 50180, Thailand}
\email{nicha@narit.or.th}

\author[0000-0002-8512-1404]{Takahiro Morishita}
\affiliation{Tohoku University, Sendai, Miyagi, Japan}
\email{morishita@astr.tohoku.ac.jp}

\author[0000-0001-9391-305X]{Michele Trenti}
\affiliation{School of Physics, The University of Melbourne, VIC 3010, Australia}
\email{michele.trenti@unimelb.edu.au}

\begin{abstract}
 
Ultra-cool dwarf stars (UCDs) are a common source of contamination in high-redshift galaxy searches because both populations have red colors and these early-forming galaxies can have sizes that are difficult to resolve even with space telescopes. We develop \textbf{F}oreground \textbf{C}ontamination \textbf{E}valuator of \textbf{N}earby dwarf stars in high-\textbf{Z} photometrically selected \textbf{O}bjects (FC-ENZO), a code that predicts magnitude-dependent UCD contamination for a given survey setup and galaxy selection method. FC-ENZO combines UCD number-density models, synthetic spectral libraries, and photometric selection criteria to estimate contaminant populations. We find that the predicted contamination is sensitive to the adopted UCD spectral model, stellar metallicity, and Galactic distribution models, with the Galactic distribution model producing variations of up to several factors. The Elf Owl models predict higher contamination than the Bobcat models, primarily because of their broader coverage of T and Y dwarf spectra. The dominant contaminants are late L to early Y-type UCDs, which are most commonly misclassified as galaxies at $z \sim 8$. For surveys with similar areas, broad-band observations from different space telescopes produce comparable numbers of contaminants at a given redshift, but contaminants are concentrated near the survey limiting magnitude, resulting in higher bright-end contamination fractions for shallower surveys. Adding medium-band filters can reduce the UCD contamination rate at $z\sim8$ by a factor of a few. At $z\gtrsim30$, F356W dropout selections can suffer substantial contamination from Y dwarfs. FC-ENZO provides a practical tool for survey design and for identifying fields suitable for spectroscopic follow-up.


\end{abstract}



\section{Introduction} 
High-redshift galaxy candidates are commonly identified using color–color selection techniques targeting Lyman-break galaxies (LBGs), 
photometric redshift estimates, or a combination of both methods. The former approach, often referred to as the Lyman-break galaxy selection technique \citep[e.g.,][]{Steidel_1999}, has been widely used because it requires observations in only a limited number of filters. In contrast, photometric redshift estimation has gained prominence over the past decade, as it exploits information from all observed bands, thereby maximizing the scientific return from multiwavelength datasets.
Photometric redshift estimation can be applied in conjunction with color–color selection \citep[e.g.,][]{borsani_2022, bouwens_2015} or used as a stand-alone selection criterion \citep[e.g.,][]{heintz2024}. 

Photometrically selected candidates can be contaminated by two types of objects: low-redshift galaxies and ultra-cool dwarfs (UCDs) --- stellar and substellar objects of spectral types M, L, T, and Y with masses $M \lesssim 0.1,M_\odot$ and effective temperatures $T_\mathrm{eff} \lesssim 3000$,K \citep{Kirkpatrick2005}. Contamination from low-redshift galaxies arises because their Balmer-breaks can produces spectral features that resemble Ly$\alpha$ breaks of high-redshift galaxies. Stellar contamination occurs because, at faint magnitudes, point sources are often difficult to distinguish morphologically from compact extended objects, especially when the SourceExtractor stellarity parameter \texttt{CLASS\_STAR} is used to quantify the likelihood that an object is point-like \citep{Finkelstein2015c,Morishita2021}. In addition, UCDs exhibit colors similar to those of high-redshift galaxies. In particular, M-type stars primarily contaminate $R$- and $V$-band dropout selections \citep{Stanway2008, Pirzkal2009}, while L- and T-type dwarfs are significant contaminants in $I$- and $Y$- band dropout samples \citep{Stanway2004, Caballero2008, Oesch2013}. 

Photometric high-redshift candidates that are later spectroscopically identified as brown dwarfs are not uncommon. An example of such contamination was reported in \citet{borsani_2025}, who followed up 10 of the 26 photometrically selected $z\sim8$ candidates reported in \citet{borsani_2022} using JWST/NIRSpec. These sources were originally selected from the Brightest of Reionizing Galaxies (BoRG) survey, a pure-parallel imaging program on the Hubble Space Telescope (HST). Of the ten targets, two were confirmed to be low-redshift contaminants, and one was identified as a T2 brown dwarf. Similarly, two objects originally identified as potential $z\sim20-30$ objects based on JWST/NIRCam imaging of the Bullet Cluster were subsequently followed up and found to be Y dwarfs \citep{bradac_2026}. Deep extragalactic imaging has also been regularly used to identify foreground Galactic dwarf populations, further demonstrating the presence of brown dwarfs in fields used for high-redshift galaxy searches \citep[e.g.,][]{Ryan2011, holwerda2014, Nonino2023, Hainline2024,hainline_2024b,hainline2026_BDs}. 

An accurate estimation of contamination is crucial for at least two key areas of high-redshift galaxy studies. The first is the determination of the galaxy luminosity function (LF), which underpins a wide range of derived quantities, including the cosmic star-formation rate density\citep[e.g.,][]{Finkelstein2023,Robertson2024}, the UV luminosity density \citep[e.g.,][]{Oesch2018,Harikane2024}, and the neutral hydrogen fraction during the epoch of reionization \citep[e.g.,][]{Robertson_2015,Whitler2025}. The second is the measurement of galaxy clustering, which links high redshift galaxies to their underlying dark matter halos through quantities such as the luminosity dependence of galaxy bias \citep[e.g.,][]{Harikane2016,Dalmasso2024a,Dalmasso2024b} and overdensity measurements \citep[e.g.,][]{Sutanto2026,Kreilgaard2026}. Even a small fraction of contaminants can significantly bias measurements at the bright end of the galaxy population, where number statistics are limited. This is especially relevant for wide-area surveys including pure-parallel space telescope programs such as BoRG, ground-based surveys \citep[e.g.,][]{Bowler2012}, and the upcoming surveys with the \textit{Nancy Roman Grace Telescopes}.

In the literature involving high-redshift galaxy selection, there are two steps commonly taken to account for UCDs contamination. The first step occurs during candidate selection, where sources that are likely to be UCDs are rejected. Most studies implement this by including UCDs templates in spectral energy distribution (SED) fitting and comparing the goodness of fit ($\chi^2$) between the best-fit galaxy template and the best-fit stellar template. Candidates that are better fit by stellar templates are then excluded \citep[e.g.,][]{Bowler2012,McLure2013,bouwens_2015,Finkelstein2022,Franco2025,hainline_2024a,hainline2026_JADES}. Other studies use color--color information to exclude the localized region occupied by brown dwarfs, often in tandem with size information such as stellarity \citep[e.g.,][]{Ouchi2009,Bouwens2011,Finkelstein2015c,Morishita2018}. Studies targeting $z\gtrsim9$, or those that rely primarily on JWST data, may not include this step because UCDs contamination is generally expected to be less significant at these redshifts \citep{Oesch2013,Weibel2026}, because photometry beyond $2\,\micron$ can help mitigate UCDs contamination, or because the sources are spatially resolved with JWST \citep{Finkelstein2024,Leethochawalit2026}.

The second step is to statistically account for the probability that each final galaxy candidate is a UCDs during the LF calculation process. This step is not included in all studies in the literature. Among those that do account for it, some statistically estimate the contamination rate based on a spectroscopic subsample \citep{Steidel_1996}, resample candidates according to the probability that each source is a dwarf \citep{leethochawalit_uv_2023}, or use injection--recovery simulations of brown dwarfs, either at the catalog level \citep{Ouchi2009} or directly in the imaging data \citep{Bowler2015,bouwens_2015}.

In this paper, we follow the catalog-level injection--recovery simulation method of \citet{Ouchi2009} and develop a code to calculate contamination from UCDs in photometrically selected high-redshift galaxy samples, \textbf{F}oreground \textbf{C}ontamination \textbf{E}valuator of \textbf{N}earby dwarf stars in high-\textbf{Z} photometrically selected \textbf{O}bjects (FC-ENZO)\footnote{Source code available at
\href{https://github.com/onnalininn/Evaluator-of-Nearby-dwarf-stars-in-high-Z-photometrically-selected-Objects-ENZO}{GitHub}. The software is archived at Zenodo \citep{Innala2026FCENZO}.}, which can be conveniently applied to a wide range of observational setups and selection methods. In this code, we incorporate the most up-to-date number density models and synthetic spectral libraries of UCDs. We also investigate contamination rates across different redshifts and surveys. These results can be used to aid observation planning, statistically account for contamination, and inform prioritization of follow-up observations and survey fields.

\section{DATA and Input file} \label{sec:data_code}
\subsection{Stellar model} \label{stellar_model}

Although several tens of thousands of UCDs have been photometrically identified \citep[e.g.,][]{Ryan2005, Kakazu2010, Sorahana2019, CarneroRosell2019, Warren2021, hainline_browndwarf_2024}, spectroscopic observations are still largely limited to stellar-type UCDs \citep[e.g.,][]{Pirzkal2005, Pirzkal2009}, with only a handful of lower-temperature UCDs observed spectroscopically \citep[e.g.,][]{Masters2012, aganze_2022a, Burgasser2024_UNCOVER, Burgasser2025_newtemplate, Morrissey2026}. Although empirical spectral libraries \citep[e.g.,][]{Burgasser_spexprism} more directly represent observed UCD spectra, their coverage in wavelength, metallicity, and spectral type remains limited. We therefore use theoretical stellar atmosphere models in our work.

We adopt the recent \textbf{Sonora Elf Owl} models \citep{sonora_elfowl} as a fiducial framework. 
Elf Owl incorporates cloud-free, radiative-convective equilibrium atmospheres with vertical mixing and disequilibrium chemistry. It spans $\log g$ from 3.25–5.5 and effective temperatures from 275 K to 2400 K, corresponding to the spectral types from late-M through Y dwarfs. It also covers the metallicity from $[\mathrm{M}/\mathrm{H}]=-1.0$ to $1.0$, where we adopt the $[\mathrm{M}/\mathrm{H}]=0$ in the fiducial setup. In total, it provides approximately 43000 atmospheric model spectra over the 0.6 to 20~$\mu$m wavelength range. This broad wavelength coverage and metallicity range are particularly valuable for modeling UCD spectral energy distributions in \textit{JWST} observations, as they encompass the wavelength range probed by \textit{JWST}/NIRCam and allow us to systematically assess how assumptions about metallicity affect the predicted contamination rates.

To compute the observed flux of each UCDs at different distances, the synthetic surface flux density must be scaled by the factor $\left(\frac{R}{D}\right)^2$, where $R$ is the object’s radius and $D$ is the distance to the object. Here, we obtain radii by interpolating as a function of effective temperature and $\log g$ using values from the \textbf{Sonora Bobcat} evolutionary tracks \citep{bobcat}, an earlier generation of models preceding the Elf Owl.

We also assign a spectral type to each model object, based on its effective temperature, in order to determine its distribution within the Milky Way (Section \ref{sec:distribution_model}). We adopt the temperature–spectral type relations from \citet{Kirkpatrick_2021}, which provide classifications across the UCDs sequence from M9 to Y2. We note that certain spectral types maybe mis-labelled—specifically the L9-T4 types corresponding to temperatures in the range 1100-1300 due to the flat relation between spectral type and effective temperature (Figure 22 in \citet{Kirkpatrick_2021}), which makes it difficult to accurately identify the spectral types. 

\subsection{Stellar Distribution Model} \label{sec:distribution_model}

To model the spatial distribution of UCDs within the Milky Way, we adopt the \citet{Juric_2008} model, with parameter values constrained by \citet{aganze_2022b} as our fiducial framework. Specifically, the expected number of sources of a given spectral type, $N(\mathrm{SpT})$, within a given field of view can be expressed as:

\begin{equation}
N(\mathrm{SpT}) = \Delta\Omega \int_{d_{\min}(\mathrm{SpT})}^{d_{\max}(\mathrm{SpT})} 
\rho(r) \cdot r^2 \, dr,
\label{eq:N(SPT)}
\end{equation}

\noindent
where $\Delta\Omega$ is the solid angle of the observed field. $d_{\min}$ and $d_{\max}$ define the distance range probed for a given spectral type. The term $\rho(r)$ denotes the number density as a function of distance with contributions from the thin disk, thick disk, and halo populations. 

The thin and thick disks are modeled as exponential distributions in Galactocentric cylindrical coordinates, with density profiles given by:

\begin{equation}
\rho_{\mathrm{disk}}(r) = \rho_{\odot} \cdot \exp\left( -\frac{R - R_{\odot}}{L} \right) \cdot \exp\left( -\frac{|Z - Z_{\odot}|}{H} \right),
\label{eq:diskmodel}
\end{equation}

\noindent
where $\rho_{\odot}$ is the local space density near the Sun, $R = \sqrt{X^2 + Y^2}$ is the Galactocentric radial distance, and $Z$ is the vertical height above the Galactic plane. Following \citet{aganze_2022b}, the thin disk vertical scale heights depend on the spectral type, e.g., late-M dwarfs: $H = 249^{+48}_{-61}$~pc, L dwarfs: $H = 153^{+56}_{-30}$~pc, T dwarfs: $H = 175^{+56}_{-149}$~pc. 

The corresponding local number densities ($\rho_{\odot}$) for each spectral type are adopted from Figure~29 of \citet{Kirkpatrick_2021}. For the thick disk component, we adopt a scale height of $H_{\mathrm{thick}} = 900$~pc. The radial scale lengths of the thin and thick disks are $L_{\mathrm{thin}} = 2600$~pc and $L_{\mathrm{thick}} = 3600$~pc, respectively, following \citet{Juric_2008}. 

The halo population follows a flattened spheroidal model:

\begin{equation}
\rho_{\mathrm{halo}}(r, z) = \left( \frac{R_{\odot}}{\sqrt{r^2 + (z/q)^2}} \right)^n,
\end{equation}

\noindent
with parameters $q = 0.64$ and $n = 2.77$. The number density ratios of the halo-to-thin disk and thick disk-to-thin disk populations are assumed to be 0.25\% and 2\%, respectively \citep{Juric_2008}.

To apply this model, we convert each object’s position from Galactic coordinates $(l, b)$ and heliocentric distance $d$ to the Cartesian coordinates $(x, y, z)$. The Galactocentric position vector $\vec{r} = (X, Y, Z)$ is calculated as:

\begin{equation}
    \begin{aligned}
        X &= R_{\odot} - d \cos b \cos l \\
        Y &= -d \cos b \sin l \\
        Z &= Z_{\odot} + d \sin b
    \end{aligned}
\end{equation}

\noindent
We adopt $R_{\odot} = 8300$~pc and $Z_{\odot} = 27$~pc \citep{Gillessen_2009,Chen_2001}.

\section{Code Structure} 
\label{code_structure}

FC-ENZO is designed to estimate the number of UCD contaminants that are misclassified as high-redshift galaxies under user-specified selection techniques within a given observation field. The supported galaxy selection methods include photometric redshift estimation using EAzY, as well as user-provided color--color selection. 
The output provides an estimate of the expected number of UCD contaminants in the target field. 
We provide an overview of the user input file and the calculation structure below, with further details available in the documentation on \texttt{GitHub}.

\subsection{User Input} \label{User Input}

\paragraph{Initial Setup and Image Quality Parameters}
The initial setup requires the user to specify the field name, as well as its coordinate and field of view, output directory, and apparent magnitude bins in a reference filter over which the contamination will be calculated. In addition, image calibration parameters must be provided for each filter used in the galaxy selection criteria: \texttt{PHOTPLAM}, \texttt{PHOTFLAM}, \texttt{EXPTIME}, and the limiting magnitude (depths). These parameters vary across fields and are used to estimate the noise of the simulated UCDs (see details in Appendix \ref{apd:noise_flux_new_mag}).

\paragraph{Selection Method Settings}
\label{selection_setting}
The user can choose whether the output should simply report the numbers of UCDs in the given field or, alternatively, the number of UCDs expected to contaminate a given galaxy sample in a given field. In the latter case, users must specify the stellar model, stellar metallicity, and the galaxy selection criteria. Currently, two options (or a combination of both) options are supported: photometric redshift selection with \texttt{EAzY} and color-color selection.

\subsection{Step 1: Generate catalogs of UCDs at different Magnitude Bins} \label{sec:generate move table}

For each apparent magnitude bin in a specified detection band, we generate a catalog of all UCDs in the stellar evolutionary model spectra. We shift the UCDs so that their flux in the detection band matches the value of the bin center, with fluxes in other filters scaled proportionally. The signal-to-noise ratio (SNR) in each filter is computed by combining photon-counting (Poisson) noise with contributions from read noise, dark current, and background noise (through $5\sigma$ limiting magnitude). The noise calculation requires calibration parameters such as \texttt{PHOTFLAM}, \texttt{PHOTPLAM}, and exposure time to convert cgs fluxes into photon counts (see appendix \ref{apd:noise_flux_new_mag} for further details). Using these ``true" shifted fluxes and the SNR estimates, we then generate the “observed” fluxes assuming a Gaussian noise.

\subsection{Step 2: Misclassification Fractions} \label{sec: misclassification}
After generating the ``observed" flux catalog, we apply galaxy selection criteria to evaluate the misclassification fraction, i.e., the fraction of UCDs in each spectral class that satisfy the galaxy selection criteria. The implemented options are (1) photometric redshift estimation using EAzY, and (2) the color–color selection method. The result from EAzY is typically used in two ways: first, by comparing the relative goodness-of-fit between stellar and galaxy templates, and second, by applying constraints on the redshift range and redshift probability distribution $p(z)$, often in combination with color–color criteria. The color–color selection uses two colors, usually constructed from magnitudes in three filters. The criteria require (i) a strong flux decrement due to absorption blueward of Ly$\alpha$, and (ii) the object not to have an excessively red continuum slope, to suppress contamination from low-redshift dusty or passive galaxies. These criteria are further supplemented by the requirement of non-detections in filters bluer than Ly$\alpha$ and a minimum signal-to-noise ratio (SNR) detection in the UV continuum. Based on the selection criteria, we calculate the misclassification fractions, e.g. the fraction of mock UCDs of each stellar type that satisfies all galaxy selection criteria.
We use the standard binomial sampling uncertainty to estimate the uncertainties of these misclassification fractions.

We note that morphology-based criteria, such as \texttt{CLASS\_STAR} from SourceExtractor, are not included since our analysis is performed at the catalog level. Moreover, the magnitude ranges in which we assess contamination typically correspond to regimes where extended sources cannot be reliably distinguished from point sources (e.g., $m\gtrsim25$ in HST data and $m\gtrsim28$ in JWST data; see \citet{Morishita_2020,Holwerda2024}). Morphology-based criteria currently used in high-$z$ galaxy selection are primarily intended to reject extended low-redshift interlopers \citep[e.g.,][]{Holwerda2020} or to complement SED fitting with stellar templates for further candidate screening \citep[e.g.,][]{bouwens_2015,Morishita2018}. Therefore, the lack of a morphology criterion is not expected to significantly affect our estimates of UCD contamination.

\subsection{Step 3: Number Density of UCDs in the observed field} \label{sec: number density}
In this step, we estimate the number of UCDs from late-M to early-Y types at each magnitude bin in the observed field. We adopt the UCDs model distribution from \citet{aganze_2022b} as described in Section \ref{sec:distribution_model}. Specficically, we use Equation \ref{eq:N(SPT)} to compute the expected number of UCDs of spectral type SpT type in the magnitude bin $m_x$, denoted $N(SpT,m_x)$. We conservatively determine the minimum and maximum line-of-sight distances, $d_{\min}$ and $d_{\max}$, by accounting for both the intrinsic scatter in UCDs luminosities and the width of the magnitude bin. Specifically, $d_{\min}$ is defined as the lower $1\sigma$ limit of the distances obtained by shifting UCDs of a given spectral type to $m_{\mathrm{bin}} + 0.5\Delta m$ mag. Similarly, $d_{\max}$ is defined as the upper $1\sigma$ limit of the distances obtained by shifting the UCDs to $m_{\mathrm{bin}} - 0.5\Delta m$ mag.

To account for uncertainties in the stellar distribution models, we sample the local density $\rho_\odot$ and scale height $H$ parameters 100 times, assuming a gaussian distribution. The reported number densities are the median and $1\sigma$ values from this Monte Carlo sampling.

\subsection{Step 4: Contamination of UCDs}
Finally, we combine the misclassification fractions from Section~\ref{sec: misclassification} with the UCDs number densities from Section~\ref{sec: number density} to estimate the expected number of UCDs that mimic high-redshift galaxies in the observed field. The final results include the predicted number of contaminants as a function of magnitude, along with plots illustrating the misclassification trends. We note that the reported uncertainties capture only the uncertainty in the mean within the assumed model. They do not account for systematic uncertainties arising from different model assumptions (see Section~\ref{sec:contamrate_and_modelparam} and Appendix~\ref{apd:limitation}) or statistical sampling uncertainties, such as Poisson or binomial noise. We recommend that users account for these sampling uncertainties when applying these predictions to their science analyses.

\section{Results}
We first present the results of testing the code on the follow-up fields from \citet{borsani_2025}. We then examine the robustness of the predicted contamination rates under different evolutionary and number density models. Finally, we discuss strategies to mitigate UCDs contamination on the HST observations.

\subsection{Demonstrating Our Pipeline for High-Redshift Follow-Up Prioritization} \label{sec:showcase}

We first check whether our contaminant estimation could help flagging the fields with possible high contamination rates from UCDs in real observations. \citet{borsani_2022} identified 26 $z\sim8$  high-confidence candidates from 288 HST parallel fields compiled in the (Super)Brightest of Reionizing Galxies (SuperBoRG) data set \citep{Morishita2021}. 
For each final candidate, they calculated the probability that the source is a brown dwarf based on the $\chi^2$ comparison, adopting a uniform brown dwarf number density across the extragalactic sky as the prior. 
Ten of the highest-priority candidates were followed-up with NIRSpec/JWST(GO 1747, \citet{borsani_2025}). Of these, seven were confirmed as high-$z$ galaxies, two were $z\sim2$ interlopers and one was confirmed to be a T2-type UCDs. Here we will apply our code to the fields of the ten $z\sim8$ candidates and see if the priorities would be altered.

We run \texttt{FC-ENZO} using the fiducial setting i.e., using the Milky Way distribution model from \citet{aganze_2022b} and the \texttt{Elf Owl} stellar model. We implement the same $z\sim8$ selection criteria as \citet{borsani_2022}, including the $Y_{098}/Y_{105}-J_{125}$ and $J_{125}-H_{160}$ color--color cuts, photometric redshift selection with \texttt{EAzY} \citep{EAZY2008} using the V1.3 galaxy template set, and rejection of sources better fit by brown dwarf templates from the \texttt{SpeX Prism Library} \citep{Burgasser_spexprism}. The best-fit photometric redshift is required to satisfy $7.5 \leq z_p \leq 8.5$ with $P(z>6.5) > 0.7$. For each field, we model the UCD contaminants over the magnitude range $24.0 < H_{160} < 26.75$, corresponding to the candidate magnitudes reported by \citet{borsani_2022}.

The results are shown in Table~\ref{table: EO_contamination}, highlighting the importance of taking into account the depths of the image and the selection method when quantifying UCD contaminants. The final column shows the estimated number of UCDs that would be selected as $z\sim8$ galaxies. The middle three columns present the number densities at different stages of the calculation: the number density of UCDs within $\text{F160W}\in[24,26.75]$, the density preferred by galaxy templates over stellar templates, and the density that satisfies the color-color and photometric redshift selection criteria. The field with confirmed UCDs contaminant, BORG-0314-6712, is highlighted in red. Although this field does not have the highest intrinsic UCDs number density according to the Milky Way model, it is predicted to have the highest expected number of UCD contaminants, $\bar{N}_{cont}\approx 0.15 \pm 0.02$ per field, once the combined effects of filter selection and image depths are taken into account.

\begin{figure*}
    \centering
    \includegraphics[width=0.95\linewidth]{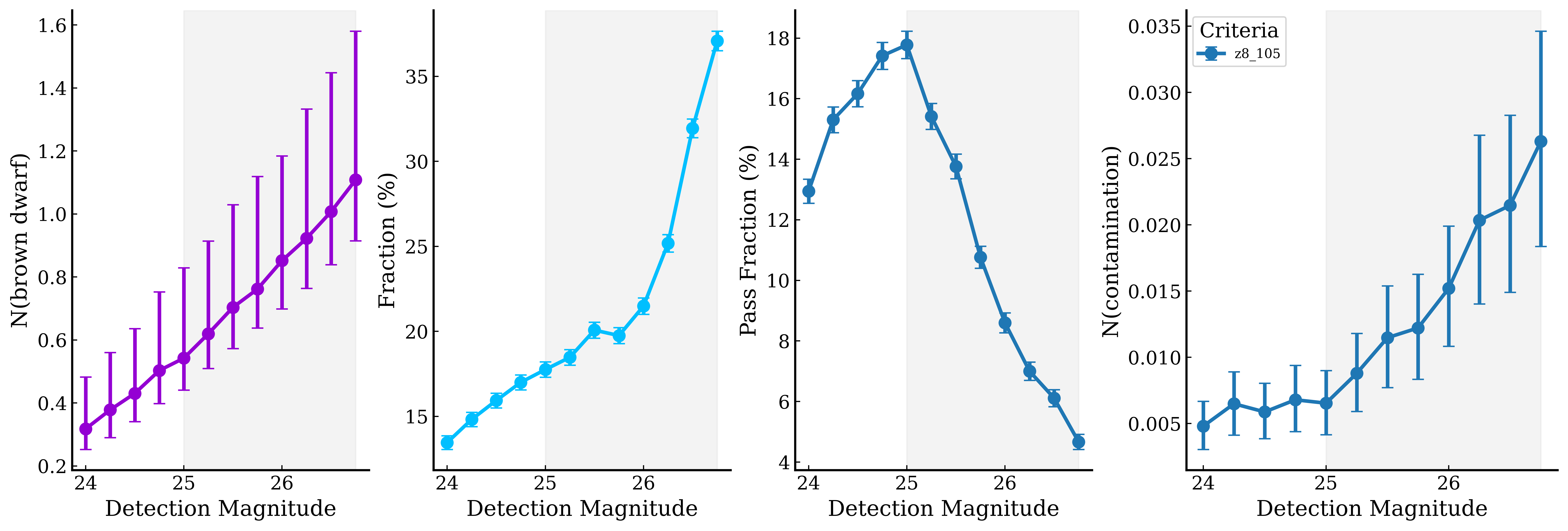}
    \caption{Example output from FC-ENZO using the fiducial model setup for the BORG-0314-6712 field with the $z\sim8$ galaxy selection criteria from \citet{borsani_2022}. The first panel shows the total number density of UCDs of all spectral types as a function of F160W magnitude, which depends on the sky position of the field. The second panel shows the fraction of UCDs for which the galaxy template is preferred over the stellar template based on a $\chi^2$ comparison. The third panel shows the fraction of UCDs that satisfy the remaining $z\sim8$ galaxy selection criteria, including the NIR color--color cuts, signal-to-noise ratio requirements in different bands, and photometric redshift criteria. The final panel shows the expected number of contaminants in $z\sim8$ sample in the field as a function of magnitude. The gray shaded region indicates the magnitude range over which stars cannot be reliably distinguished from extended sources in HST imaging using the stellarity parameter \citep{Morishita2021}.
    }
    \label{fig: 0314-6712_EO_contaminate}
\end{figure*}

Figure~\ref{fig: 0314-6712_EO_contaminate} shows the results as a function of F160W magnitude for the BORG-0314-6712 field, corresponding to each column of Table~\ref{table: EO_contamination}. The first panel indicates that the number of UCDs increases toward fainter magnitudes, as expected from the larger integration volume at greater distances from the Sun. The second panel demonstrates that the $\chi^2$-based template comparison becomes less reliable near the limiting magnitude (the $5\sigma$ limiting magnitude in F160W for this field is $\sim$27.09~mag). In the faintest bin, nearly half of the synthetic stars preferentially select galaxy templates over brown dwarf templates. In contrast, the third panel demonstrates that the galaxy selection criteria—combining color--color cuts and photometric redshift constraints—remain robust against dwarf contamination. Fewer than $\sim$20\% of synthetic dwarfs are misclassified as galaxies. Near the detection limit, this fraction decreases to a few percent, as many sources do not satisfy the detection criteria. 
The final panel shows the expected number of contaminants in the HST field as a function of magnitude. In general, the contamination is highest near the limiting magnitude and in the magnitude range where the stellarity parameter cannot be used to reliably distinguish a point source from an extended source \citep[grey shade;][]{Morishita2021}.



\begin{table*}
\footnotesize
\renewcommand{\arraystretch}{1.2}
\setlength{\tabcolsep}{2.5pt}
\begin{center}
\caption{Contaminant Numbers in Elf Owl: MH = 0.0}
\begin{tabularx}{\textwidth}{@{}
>{\raggedright\arraybackslash}p{2.7cm}
>{\centering\arraybackslash}p{3cm}
>{\centering\arraybackslash}p{3.5cm}
>{\centering\arraybackslash}p{3.5cm} |
>{\centering\arraybackslash}p{3.5cm} @{}}
\hline\hline
\textbf{JWST Field} &
\textbf{Number Density (per HST field)} &
\textbf{Objects preferred as galaxies based on $\chi^2$ (per HST field)} &
\textbf{Objects selected with photoz+color--color selection (per HST field)} &
\textbf{Final contaminants (per HST field)} \\
\hline
BORG-0037-3337 & $4.05 \pm 0.35$ & $1.18 \pm 0.11$ & $0.45 \pm 0.04$ & $0.08 \pm 0.01$ \\
\textcolor{red}{BORG-0314-6712\textsuperscript{\textdagger}} & $8.12 \pm 0.81$ & $1.89 \pm 0.21$ & $0.88 \pm 0.09$ & $0.15 \pm 0.02$ \\
BORG-0409-5317 & $6.41 \pm 0.63$ & $1.53 \pm 0.19$ & $0.02 \pm 0.00$ & $0.01 \pm 0.00$ \\
BORG-0440-5244\textasteriskcentered & $7.21 \pm 0.76$ & $5.50 \pm 0.58$ & $0.00 \pm 0.00$ & $0.00 \pm 0.00$ \\
BORG-0853+0309 & $8.79 \pm 1.03$ & $2.20 \pm 0.32$ & $1.14 \pm 0.14$ & $0.15 \pm 0.02$ \\
BORG-0859+4114 & $4.88 \pm 0.54$ & $1.53 \pm 0.20$ & $0.41 \pm 0.06$ & $0.03 \pm 0.00$ \\
BORG-0955+4528 & $3.98 \pm 0.41$ & $1.26 \pm 0.15$ & $0.40 \pm 0.05$ & $0.06 \pm 0.01$ \\
BORG-1033+5051\textasteriskcentered & $3.79 \pm 0.38$ & $0.67 \pm 0.08$ & $0.00 \pm 0.00$ & $0.00 \pm 0.00$ \\
BORG-1437+5044 & $4.93 \pm 0.47$ & $0.83 \pm 0.10$ & $0.00 \pm 0.00$ & $0.00 \pm 0.00$ \\
BORG-2203+1851\textasteriskcentered & $16.42 \pm 1.89$ & $4.69 \pm 0.67$ & $0.00 \pm 0.00$ & $0.00 \pm 0.00$ \\
\hline
\end{tabularx}
\label{table: EO_contamination}
\vspace{0.5em}
\noindent\parbox{\textwidth}
{\footnotesize
\textit{Note:} The first column lists the HST fields that were followed-up with JWST/NIRSpec by \citet{borsani_2025}. 
The second column shows the number of expected UCDs in each field under our fiducial assumptions. 
The third column gives the expected number of UCDs classified as galaxies based on whether the best-fit EAzY galaxy template yields a lower $\chi^2$ value than the best-fit stellar template. 
The fourth column lists the number of expected UCDs that satisfy both the color–color and photometric-redshift selection criteria for $z\sim8$ galaxies. 
All selections use the F105W-dropout criterion, except those marked with * which use the F098W-dropout criterion (no F105W available). 
The final column shows the expected number of contaminants in each field, which combines the number density in column 2 and the criteria in columns 3 and 4. The field containing a confirmed brown-dwarf contaminant (BORG-0314-6712) is marked in red and indicated with a \textsuperscript{\textdagger} symbol.}
\end{center}
\end{table*}

\subsection{Contamination rates and Model Parameters}
\label{sec:contamrate_and_modelparam}
In this section, we examine the robustness of our contamination estimates and assess their sensitivity to the underlying model assumptions.

\subsubsection{Dependence on Stellar Evolution Models}
Here, we repeat the analysis described above, but now using the \textbf{Sonora Bobcat} model to represent UCDs.
The Sonora Bobcat suite has been widely adopted in the literature \citep[e.g.,][]{marley_bobcat_2021,Lacy_2023}, and is an earlier generation of the Elf Owl suite. The primary difference is that the Sonora Bobcat models assume chemical equilibrium, while Elf Owl does not. The grids span surface gravities of $3.25 \leq \log g~(\mathrm{cgs}) \leq 5.5$, effective temperatures from $200~\mathrm{K}$ to $2400~\mathrm{K}$, and metallicities of $[\mathrm{M}/\mathrm{H}] = -0.5$, $0.0$, and $+0.5$. In total, the library contains 850 spectra, each covering wavelengths from 0.6 to 20~$\mu$m.

\begin{table*}
\footnotesize
\renewcommand{\arraystretch}{1.2}
\setlength{\tabcolsep}{2pt}
\begin{center}
\caption{Summary of contamination statistics for all BORG fields.}
\begin{tabularx}{\textwidth}{@{}
>{\raggedright\arraybackslash}p{2.7cm}
>{\centering\arraybackslash}m{1.8cm} 
>{\centering\arraybackslash}m{1.8cm} |
>{\centering\arraybackslash}m{1.9cm} 
>{\centering\arraybackslash}m{1.9cm} |
>{\centering\arraybackslash}m{1.9cm} 
>{\centering\arraybackslash}m{1.9cm}
>{\centering\arraybackslash}m{1.5cm}
@{}}
\hline\hline
Field &
\multicolumn{2}{c|}{\parbox[c][2.5em][c]{4.3cm}{\centering Percentage of objects preferred as galaxies based on $\chi^2$}} &
\multicolumn{2}{c|}{\parbox[c][2.5em][c]{4.5cm}{\centering Percentage of objects selected with photoz+color--color selection}} &
\multicolumn{3}{c}{\parbox[c][2.5em][c]{5.3cm}{\centering Final contamination per HST field }} \\
\hline
\rule{0pt}{1em} & Elf Owl & Bobcat & Elf Owl & Bobcat & Elf Owl & Bobcat  & Ratio\\
\hline
BORG-0037-3337 & $1.18 \pm 0.11$ & $1.58 \pm 0.15$ & $0.45 \pm 0.04$ & $0.12 \pm 0.01$ & $0.08 \pm 0.01$ & $0.02 \pm 0.01$ & $4.00 \pm 2.06$ \\
\textcolor{red}{BORG-0314-6712\textsuperscript{\textdagger}} & $1.89 \pm 0.21$ & $2.64 \pm 0.28$ & $0.88 \pm 0.09$ & $0.14 \pm 0.02$ & $0.15 \pm 0.02$ & $0.02 \pm 0.01$ & $7.50 \pm 3.88$ \\
BORG-0409-5317 & $1.53 \pm 0.19$ & $3.39 \pm 0.35$ & $1.6 \pm 0.3\times 10^{-2}$ & $0.00 \pm 0.00$ & $1.1 \pm 0.2\times 10^{-2}$ & $0.00 \pm 0.00$ & -- \\
BORG-0440-5244* & $5.50 \pm 0.58$ & $5.85 \pm 0.59$ & $1.2 \pm 0.4\times 10^{-3}$ & $0.00 \pm 0.00$ & $1.2 \pm 0.4\times 10^{-3}$ & $0.00 \pm 0.00$ & -- \\
BORG-0853+0309 & $2.20 \pm 0.32$ & $2.82 \pm 0.39$ & $1.14 \pm 0.14$ & $0.37 \pm 0.05$ & $0.15 \pm 0.02$ & $0.02 \pm 0.01$ & $7.50 \pm 3.88$ \\
BORG-0859+4114 & $1.53 \pm 0.20$ & $1.82 \pm 0.21$ & $0.41 \pm 0.06$ & $0.03 \pm 0.01$ & $3.1 \pm 0.4\times 10^{-2}$ & $1.4 \pm 1.1\times 10^{-3}$ & $2.21 \pm 1.75 $\\
BORG-0955+4528 & $1.26 \pm 0.15$ & $1.56 \pm 0.18$ & $0.40 \pm 0.05$ & $0.12 \pm 0.02$ & $0.06 \pm 0.01$ & $1.8 \pm 0.5\times 10^{-2}$ & $3.00 \pm 0.50$ \\
BORG-1033+5051* & $0.67 \pm 0.08$ & $1.16 \pm 0.12$ & $1.9 \pm 0.5\times 10^{-3}$ & $1.0 \pm 1.0\times 10^{-3}$ & $1.9 \pm 0.5\times 10^{-3}$ & $1.0 \pm 1.0\times 10^{-3}$ & -- \\
BORG-1437+5044 & $0.83 \pm 0.10$ & $2.12 \pm 0.20$ & $3.0 \pm 0.8\times 10^{-3}$ & $0.00 \pm 0.00$ & $2.1 \pm 0.6\times 10^{-3}$ & $0.00 \pm 0.00$ & -- \\
BORG-2203+1851* & $4.69 \pm 0.67$ & $8.41 \pm 1.01$ & $0.00 \pm 0.00$ & $0.00 \pm 0.00$ & $0.00 \pm 0.00$ & $0.00 \pm 0.00$ & -- \\
\hline
\end{tabularx}
\label{table:compare_model_mh0}
\vspace{0.5em}
\noindent\parbox{\textwidth}
{\footnotesize
\textit{Note:} This table presents the fractional percentages of objects in each field. The last column lists the contamination in each field, calculated by multiplying the fraction by the number density from Table~\ref{table: EO_contamination}. Values marked with * correspond to results derived using the $z8\_098$ criterion, as $z8\_105$ data were not available. Each column value has been scaled by its respective fixed power of ten, as indicated in the header units. The field containing a confirmed brown-dwarf contaminant (BORG-0314-6712) is marked in red and indicated with a \textsuperscript{\textdagger} symbol.}
\end{center}
\end{table*} 

Here we compare the contamination results using solar-metallicity UCDs ($[\mathrm{M}/\mathrm{H}] = 0$) from both model suites on the same ten BoRG fields as in Section \ref{sec:showcase}. All other parameters and procedures remain consistent with the previous setup. We show the difference in the contamination results in Table~\ref{table:compare_model_mh0}. The \texttt{Sonora Elf Owl} model gives contamination rates up to seven times higher than the \texttt{Sonora Bobcat} model in the fields with higher contamination rates. In the field with confirmed contaminant, the \texttt{Bobcat} model predicts approximately $0.02$ UCD contaminants per field for the $z\sim8$ selection. Four out of ten fields show zero UCD contaminants per field.

The discrepancy in contamination levels can be attributed to two key differences between the two model grids. First is the number of stars in the model grids. The \texttt{Elf Owl} models contain far more spectra at solar metallicity (7,200 targets) compared to the \texttt{Bobcat} model (429 targets), making Bobcat less sensitive in the low-contamination regime, yielding zero contaminant. Second, Bobcat assumes chemical equilibrium, whereas Elf Owl incorporates disequilibrium chemistry. As a result, some \texttt{Elf Owl} spectra exhibit enhanced features beyond 1~$\mu$m \citep{PICASO_2023}, producing slightly redder SEDs. This difference translates into distinct distributions in the best-fit redshifts ($z_p$) derived with EAzY. In general, both evolution grids produce a bimodal distribution: one at 1.7 and another at $z\sim7$--8. However, the Elf owl grid yields a more complex structure at higher redshift, with two peaks at $z \sim 7.4$ and $z \sim 7.8$, while \texttt{Bobcat} produces a single peak at $z \sim 7.3$. Because the photometric-redshift selection in \citet{borsani_2022} requires $z_p>7.5$, a larger fraction of \texttt{Elf Owl} UCDs satisfy the galaxy selection criteria compared to those from the \texttt{Bobcat} grid. 

Given that the Elf Owl grid includes a larger number of spectra and incorporates more up-to-date physics, we adopt the \texttt{Elf Owl} grid as our fiducial model grid. All subsequent analyzes in this work are mainly based on the Elf Owl models.

\subsubsection{Dependence on Stellar Metallicities} \label{compare_mh}
In previous sections, we have used theoretical spectra assuming solar metallicity, $[\mathrm{M}/\mathrm{H}] = 0$. In reality, Milky Way stars span a range of metallicities. Halo stars are metal poor with $[\mathrm{Fe}/\mathrm{H}]\lesssim-1.2$ \citep[e.g.,][]{Lee2017, Conroy2019}. In the solar neighborhood, most stars cluster around solar metallicity, with $[\mathrm{Fe}/\mathrm{H}]$ ranging from approximately $-0.5$ to $0.5$ \citep{Haywood_mh2001, mh_in_milkyway}. The disk structure is further complicated by corrugations and gradients in both radial and vertical directions \citep[e.g.,][]{Chiappini1997, bland-hawthorn_galah_2019, Dietz2020}.
 We explore how the contamination rate changes when adopting different fixed stellar metallicities, varying $[\mathrm{M}/\mathrm{H}]$ between runs. We used all available metallicities in the Elf Owl models, including $[\mathrm{M}/\mathrm{H}]=\pm1.0$, $+0.7$, $\pm0.5$, and $0.0$. 
 
\begin{figure}
    \centering
    \includegraphics[width=0.95\linewidth]{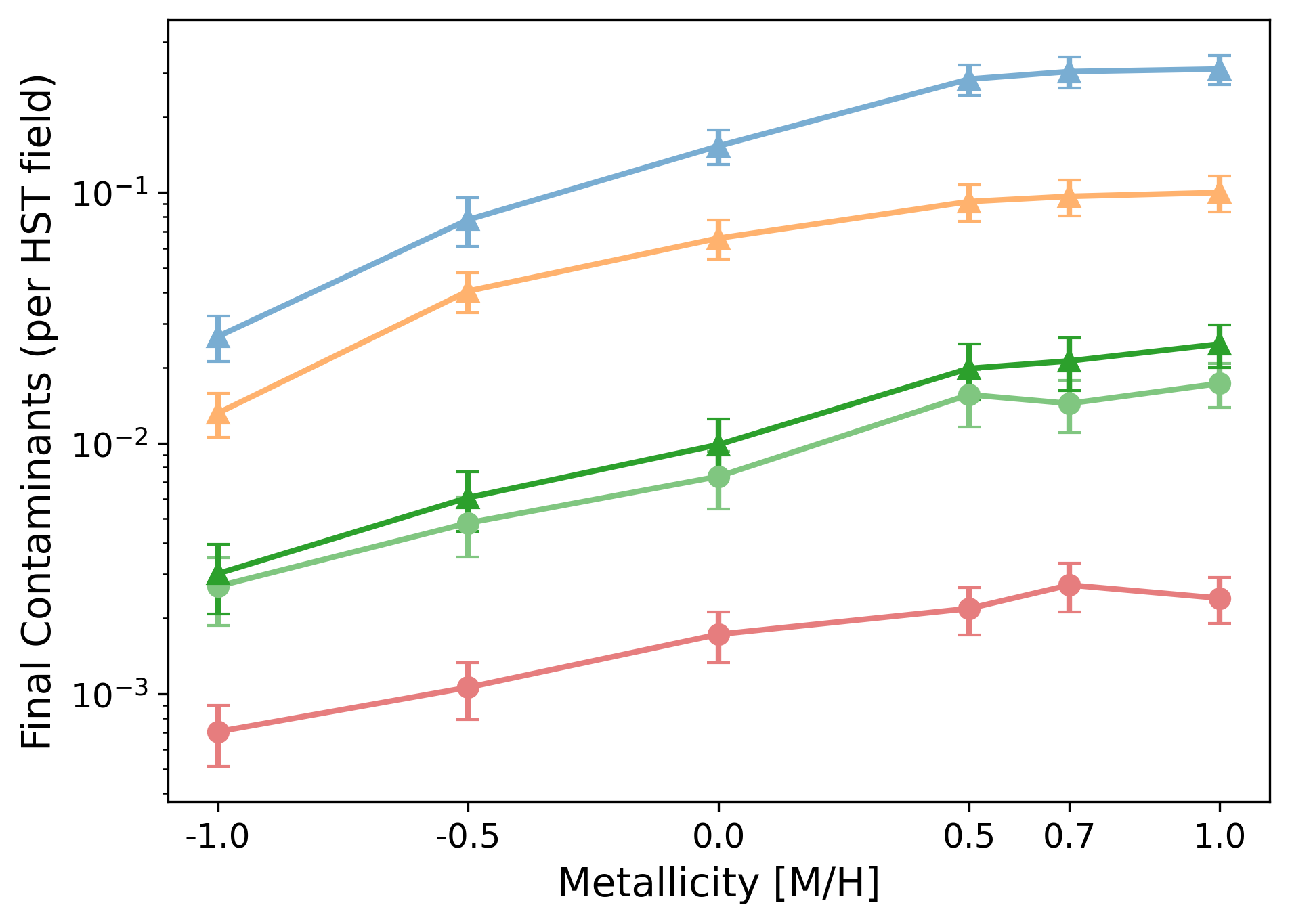}
    \label{fig:metallicity_all}
    \caption{Total contamination as a function of metallicity.
    The contamination numbers are summed over the magnitude range 24.0--26.75 (in 0.25 magnitude bins) to represent the total contamination at each metallicity and are calculated for four different BoRG fields for $z\sim8$ galaxies. The \textcolor{blue}{blue}, \textcolor{orange}{orange}, and \textcolor{teal}{green} lines marked with triangle data points are from the BORG-0314-6712, BORG-0955+4528, and BORG-0409-5317 respectively, using the $F105W$ dropout criteria. The \textcolor{teal}{light green} and \textcolor{red}{red} lines marked with circle data points are from the BORG-0409-5317 and BORG-1033+5051 fields respectively, using the $F098M$ dropout criteria.
    }
    \label{fig:compare_mh}
\end{figure}

 We present the results on five randomly-selected BoRG fields with $z\sim8$ galaxy selection criteria in Figure~\ref{fig:compare_mh}. The figure shows the total contamination numbers, summed over the magnitude range 24.0--26.75, as a function of metallicity. A clear trend is observed: the contamination for the $z\sim8$ galaxy selection strongly increases with metallicity in all fields. Generally, the contamination rate at $[\mathrm{M}/\mathrm{H}]=+1.0$ is 0.5--1~dex higher than that at $[\mathrm{M}/\mathrm{H}]=-1.0$. 
 
 We note that each metallicity in the \texttt{Bobcat} and the \texttt{Elf Owl} is associated with a range of C/O ratios, spanning 0.5--1.5 times and 0.5--2.5 times the solar value, respectively. The latter range encompasses the full distribution of the C/O ratios observed in nearby M dwarfs, while the former covers most of the observed distribution \citep{Nakajima2016}. In our default analysis, we include all available spectra. To assess the impact of varying C/O ratios, we repeated the analysis using only spectra with a solar C/O ratio and found that the results remain unchanged within 1 $\sigma$. 

In the context of our contamination calculation, fully incorporating these chemical inhomogeneities is non-trivial. At a fixed apparent magnitude, potential UCD contaminants originate from a wide range of distances, and different spectral types probe different Galactic volumes. For example, F160W = 26 mag corresponds to an M9 dwarf located at approximately 3--10 kpc, whereas a Y0 dwarf at the same apparent magnitude would lie at 0.13 kpc from the Sun. These variations imply that a self-consistent treatment would require a detailed Galactic chemical-evolution model coupled to a 3D stellar-density distribution, which is beyond the scope of the present work. 

However, given that the vertical metallicity gradient of the Milky Way is negative, e.g., $-0.22\pm0.17$ dex kpc$^{-1}$ \citep{Kordopatis2013}, and that the metallicity distribution of stellar populations in the solar neighborhood peaks at $[\mathrm{M}/\mathrm{H}] = 0.0$ \citep{Haywood_mh2001}, we can safely assume that the UCDs in the extragalactic fields are likely to have subsolar metallicities. Therefore, the contamination rates estimated using the fiducial $[\mathrm{M}/\mathrm{H}] = 0.0$ should be conservative.
                            
\subsubsection{Dependence on Stellar Distribution Model} \label{compare_dist_model}
The main source of uncertainty in the stellar distribution model is likely the estimation of the scale heights. Our fiducial scale heights are taken from \citet{aganze_2022b}, who derived them from a sample of 164 late-M, L, and T dwarfs identified in the WISP and 3D-HST grism surveys \citep{aganze_2022a}. For late-M dwarfs, they reported a scale height of $H = 249^{+48}_{-61}$~pc. Scale height estimates based on earlier HST observations, which primarily extended to M-type stars, range from 290 to 400 pc, with uncertainties of 50--100 pc \citep[e.g.,][]{Pirzkal2009,holwerda2014,van_vledder_2016}. Although these values are somewhat higher, they remain broadly consistent with the fiducial value. A more recent estimate based on SDSS-UKIDSS data reported a similar value of $H\sim270$ pc \citep{Warren2021}.

In this section, we instead evaluate our fiducial number density estimates against the stellar distribution models of \citet{honaker_simulating_2025}. Unlike most previous estimates, which derive scale heights from stellar number counts, \citet{honaker_simulating_2025} derived their scale heights from the star formation history as a function of distance from the Galactic plane, inferred from Gaia data. Their space-density model consists of thin- and thick-disk components, each described by an exponential profile (Equation~\ref{eq:diskmodel}), without a separate halo component. The thin-disk scale heights are $113 \pm 10$~pc for L dwarfs, $123 \pm 11$~pc for T dwarfs, and $124 \pm 11$~pc for Y dwarfs, approximately $30\%$ smaller than those adopted by \citet{aganze_2022b}. The adopted values of $L_{\mathrm{thin}}$, $H_{\mathrm{thick}}$, and $L_{\mathrm{thick}}$ are identical to those in \citet{aganze_2022b}. The model places the Sun at Galactic coordinates $(x, y, z) = (0, 0, 17.7)$~pc, following \citet{mazzi_dissecting_2024}.

\begin{figure*}
    \centering
    \includegraphics[width=\linewidth]{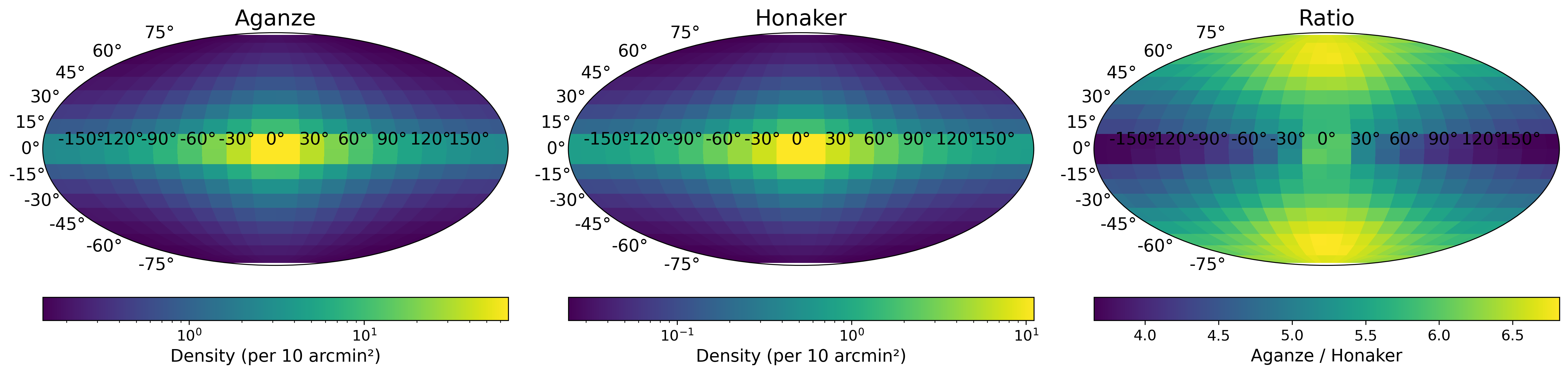}
    \caption{
    Contamination density maps in galactic coordinates using two different stellar number-density models.
    The left and middle panel show the predicted UCDs contamination density based on the MW distribution from \citet{aganze_2022b}, and \citet{honaker_simulating_2025}, respectively.
    The right panel presents the ratio between the two models, highlighting the differences in their predicted contamination patterns.
    }
    \label{fig:compare_MW_models}
\end{figure*}

Figure~\ref{fig:compare_MW_models} shows the density maps of UCD contaminants in galactic $(l,b)$ coordinate predicted by the MW models of \citet{aganze_2022b} and \citet{honaker_simulating_2025}, along with their ratio. We assume the filter set and depth of the BORG-0341-6712 field and estimate the number of contaminants over the magnitude range $24.00 \leq m \leq 26.75$. The model of \citet{aganze_2022b} predicts a higher number of contaminants than that of \citet{honaker_simulating_2025}, with differences of approximately a factor of 5--7 in typical extragalactic fields. 

These differences are primarily driven by the adopted thin-disk scale heights, which are smaller in the \citet{honaker_simulating_2025} model. At $m=26$, L to late-T dwarfs are typically located at distances of $\gtrsim500$~pc. As a result, their predicted number densities are more strongly suppressed in the \citet{honaker_simulating_2025} model because of its smaller thin-disk scale heights.

In summary, we explored the dependence of predicted UCD contaminants in $z\sim8$ galaxy selection against different assumptions in the models. We find that the stellar evolution grid can affect the final contaminant rates by up to a factor of seven and stellar metallicities by at most a factor of a few, assuming that the stars in extragalactic fields are likely solar and subsolar. Stellar distribution models can additionally alter the contamination rate by a factor of up to seven in the extragalactic fields. The absolute number of UCD contaminants in any given field remains highly sensitive to the adopted model assumptions, highlighting the need to further constrain stellar distribution models and evolutionary tracks. However, as demonstrated in Section~\ref{sec:showcase}, the relative contaminant rates between different fields and selection methods can be robustly predicted.

Additional effects not included in our fiducial model, including potential contributions from stellar substructures, such as stellar streams, spiral arms, and north--south asymmetries, as well as stellar multiplicity, are discussed in Appendix~\ref{apd:limitation}. These effects are expected to introduce additional systematic uncertainties in the predicted contamination rates, but are likely to be subdominant to the uncertainties explored in this section.

\subsection{Lessons Learned (HST Focused)}
\label{sec:HSTlesson}
\subsubsection{Contaminants in Different Redshift Selections}
In the previous sections, we focused on contaminants in the $z\sim8$ galaxy selection. In this section, we investigate ultra-cool dwarf contamination in galaxy selections at other redshift ranges. We adopted imaging data from the Cosmic Evolution Survey (COSMOS; \citealt{skelton_3d-hst_2014,Grogin_2011,Koekemoer2011}). The original mosaic covers an area of approximately $183.9,\mathrm{arcmin}^2$. However, for consistency with previous sections, we assume an effective survey area of $4.1,\mathrm{arcmin}^2$, corresponding to the field of view of a single \textit{HST} pointing.
The simulated source catalogs are constructed using the F606W, F814W, F125W, F140W, and F160W filters, with corresponding $5\sigma$ limiting magnitudes of 26.8, 26.4, 26.3, 25.7, and 26.1, respectively.

\begin{figure*}
    \centering
    
    \begin{subfigure}{0.95\linewidth}
        \centering
        \includegraphics[width=\linewidth]{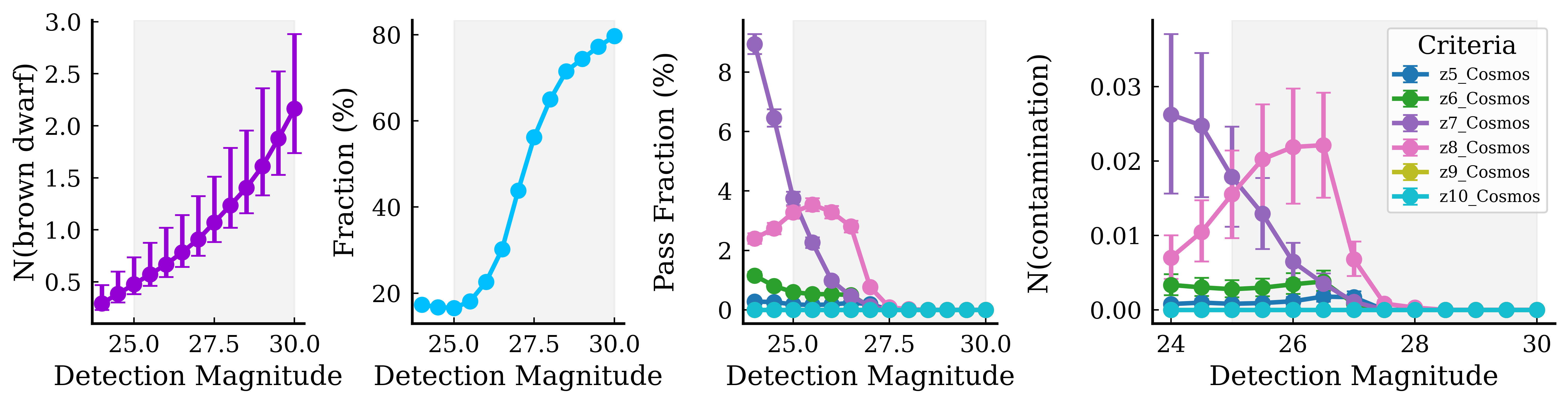}
        \caption{HST result (COSMOS)}
        \label{fig:hst_result}
    \end{subfigure}
        
    \begin{subfigure}{0.95\linewidth}
        \centering
        \includegraphics[width=\linewidth]{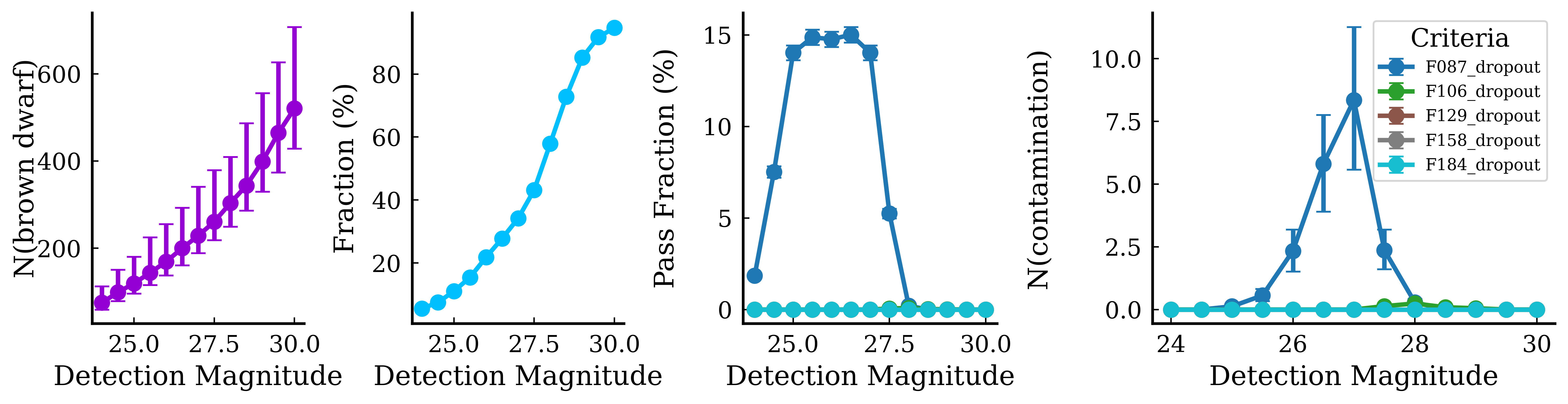}
        \caption{Roman result (Deep-tier Roman’s High-Latitude Wide-Area Survey)}
        \label{fig:roman_result}
    \end{subfigure}
    \begin{subfigure}{0.95\linewidth}
        \centering
        \includegraphics[width=\linewidth]{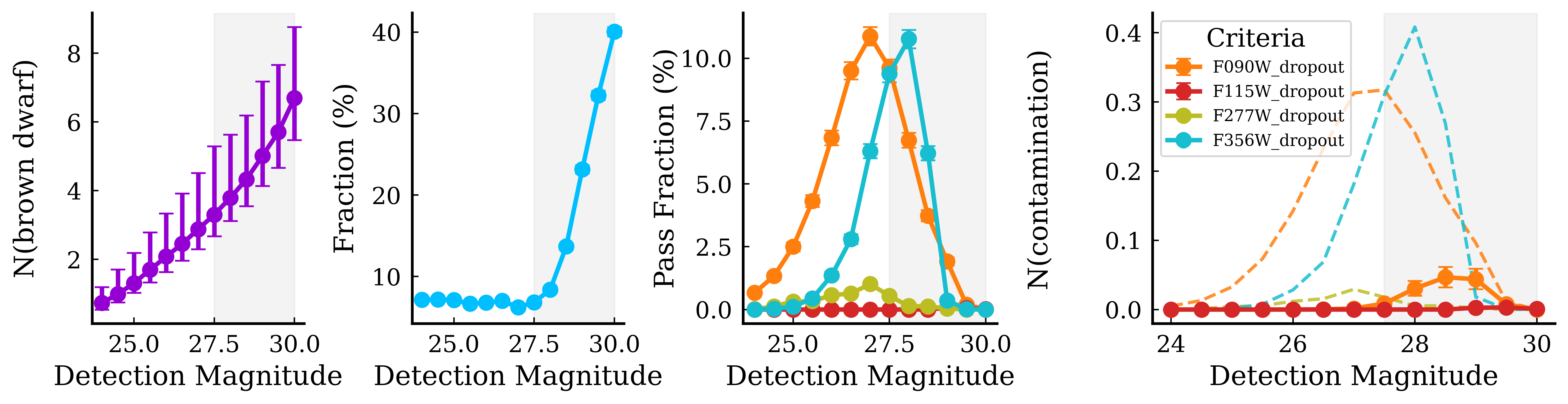}
        \caption{JWST result (BEACON Survey)}
        \label{fig:jwst_result}
    \end{subfigure}


    \begin{subfigure}{0.95\linewidth}
        \centering
        \includegraphics[width=\linewidth]{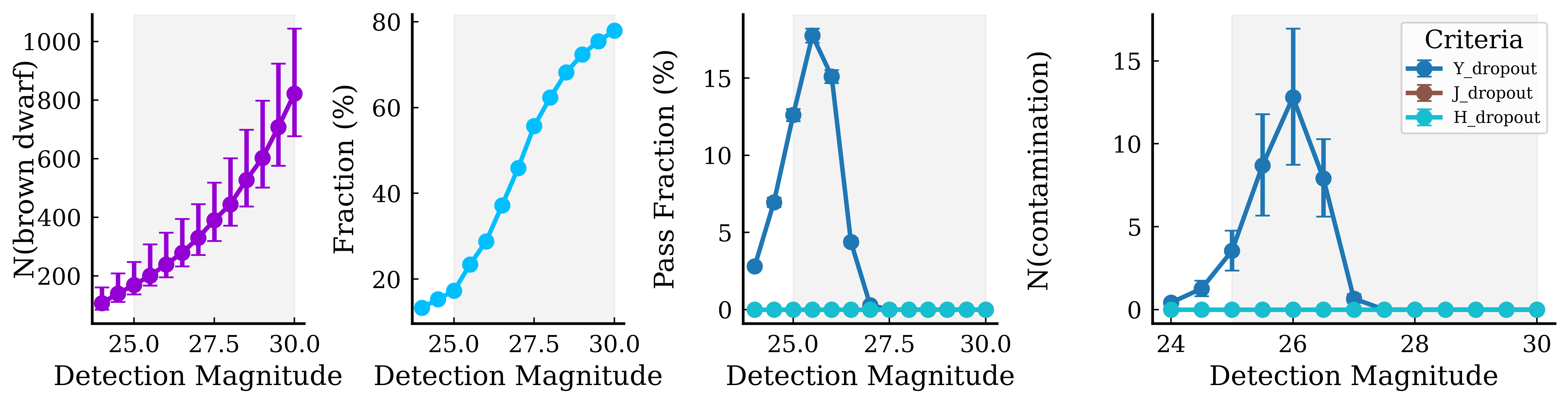}
        \caption{EUCLID result (EDFN Survey)}
        \label{fig:euclid_result}
    \end{subfigure}

    \caption{Similar to Fig.~\ref{fig: 0314-6712_EO_contaminate}, but for the 3DHST COSMOS, Roman HLWA-1000+0211, JWST BEACON-0959+0200, and Euclid EDFN fields (top to bottom). The detection magnitudes are in F160W, F158, F444W, and H, respectively. The UCD number densities (first panels) and expected contaminants (last panels) are given per telescope pointing: 4.1~arcmin$^{2}$, 0.281~deg$^{2}$, 9.7~arcmin$^{2}$, and 0.53~deg$^{2}$, respectively. The first panels show the fiducial stellar number-density model \citep{aganze_2022b}. The gray shading marks the magnitude range where stars cannot be reliably distinguished from extended sources using stellarity classification. Dashed lines in the last panel of (c) show the expected contamination without the $\chi^2$ comparison step (see text).
    }
    \label{fig:spacetelescope_result}
\end{figure*}

We follow the galaxy selection criteria from \citet{bouwens_2015} for the $z\sim5$, 6, 7, 8, and 10 samples. For the $z\sim9$ sample, we adopt the criteria from \citet{borsani_2022}. All criteria include color--color selection criteria, photometric redshift criteria, and the $\chi^2$ comparisons between galaxy templates and \texttt{SpeX} stellar templates. For the photometric redshift constraints with \texttt{EAZY}, we use the template set from \citet{Hainline2024} instead of the \texttt{EAZY\_v1.0} template set. The results are presented in Fig.~\ref{fig:hst_result}.

We find that the $z\sim8$ selection is the most susceptible to contamination from UCDs. Across all redshift ranges, both the predicted stellar number density and the fraction of synthesized stars that favor galaxy templates increase toward fainter magnitudes, as shown in the first two panels. The differences in the final contamination rates are, therefore, primarily driven by the galaxy selection criteria. 
For the $z\sim5$ and $z\sim6$ selections, the number of contaminants remains negligible at all magnitudes. For the $z\sim7$ selection, the contamination peaks at a brighter magnitude but decreases quickly to nearly zero by $m_{F160W}\sim27.5$. This is not particularly concerning, since point sources can still be morphologically distinguished at these bright magnitudes. In contrast, the contaminants in the $z\sim8$ selection peaks near $m\sim26$, which lie in the magnitude range where the morphological separation between point sources and extended sources becomes difficult. The total expected contamination from UCDs is approximately $0.1$ stars per field for both the $z\sim7$ and $z\sim8$ selections, comparable to the more highly contaminated fields in the BoRG sample, likely due to the similar choices of filters and survey depths.
 
\subsubsection{Sources of Contamination and Dropout Filters for \texorpdfstring{$z\sim8$}{z~8} Selection}

In the previous section, we showed that the $z\sim8$ selection is the most susceptible to contamination from UCDs. For \textit{HST}, two commonly used dropout selections for $z\sim8$ are the F098M- and F105W-dropout selections. Interestingly, the top three lines in Figure~\ref{fig:compare_mh} correspond to fields that use the F105W-dropout criteria, while the bottom two lines, which exhibit lower contamination rates, correspond to fields that use F098M-dropout criteria. We therefore compare the performance of these two selection criteria presented in \citet{borsani_2022}. The two selections are identical except for their color--color cuts: the F098M selection adopts $\text{F098M} - \text{F125W} > 1.75$, whereas the F105W selection uses $\text{F105W} - \text{F125W} > 0.45$.

\begin{figure*}
    \centering
    \begin{minipage}[t]{0.46\textwidth}
        \vspace{0pt} 
        \includegraphics[width=\linewidth]{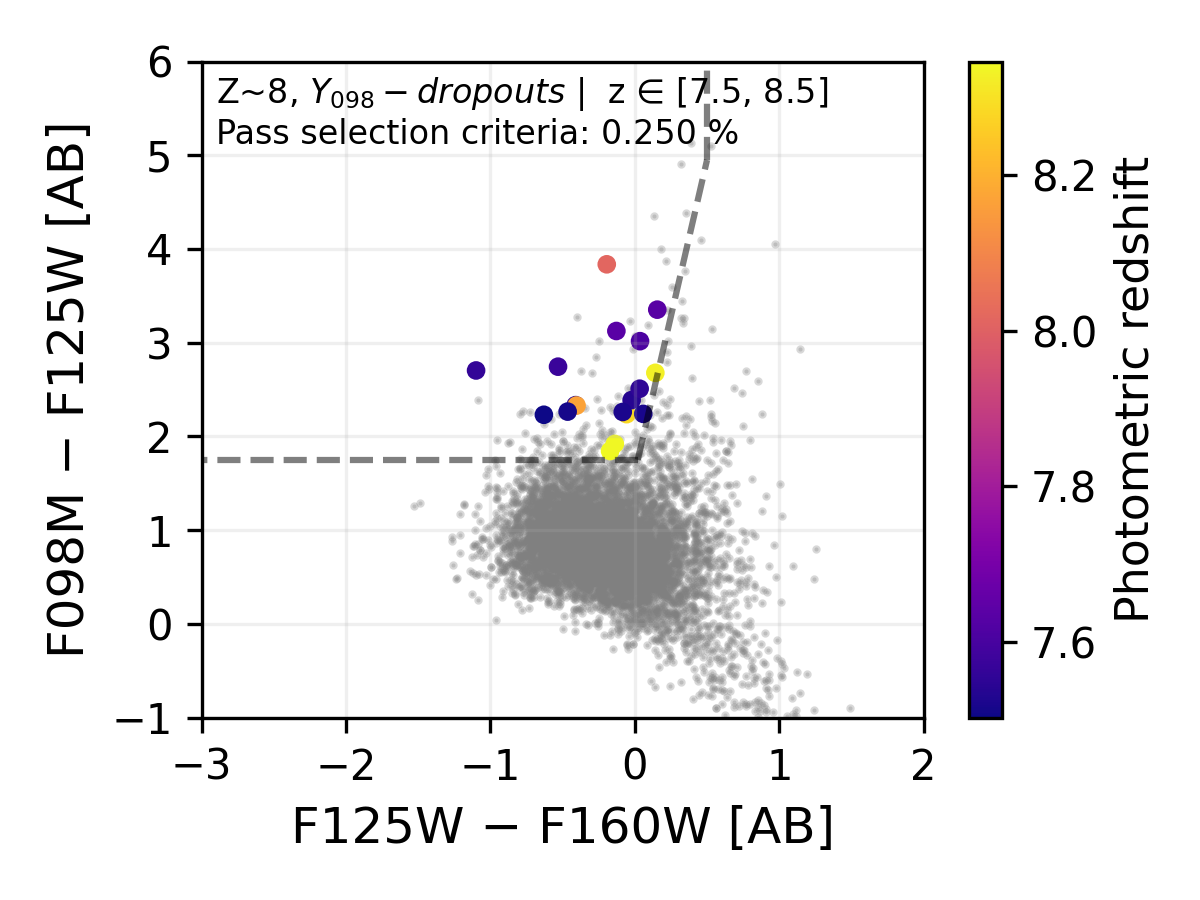}
        \includegraphics[width=\linewidth]{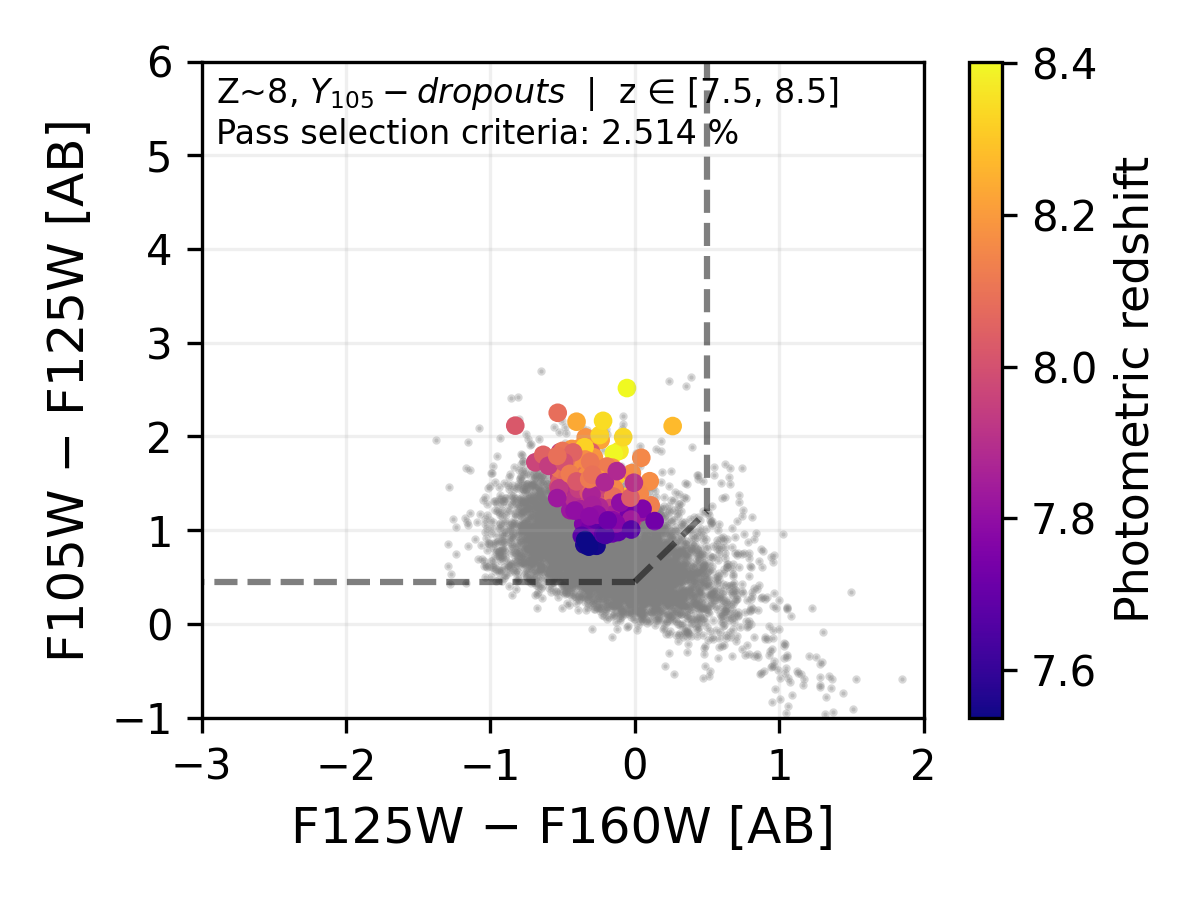}
    \end{minipage}
    \hfill
    \begin{minipage}[t]{0.43\textwidth}
        \vspace{0pt}
        \includegraphics[width=\linewidth]{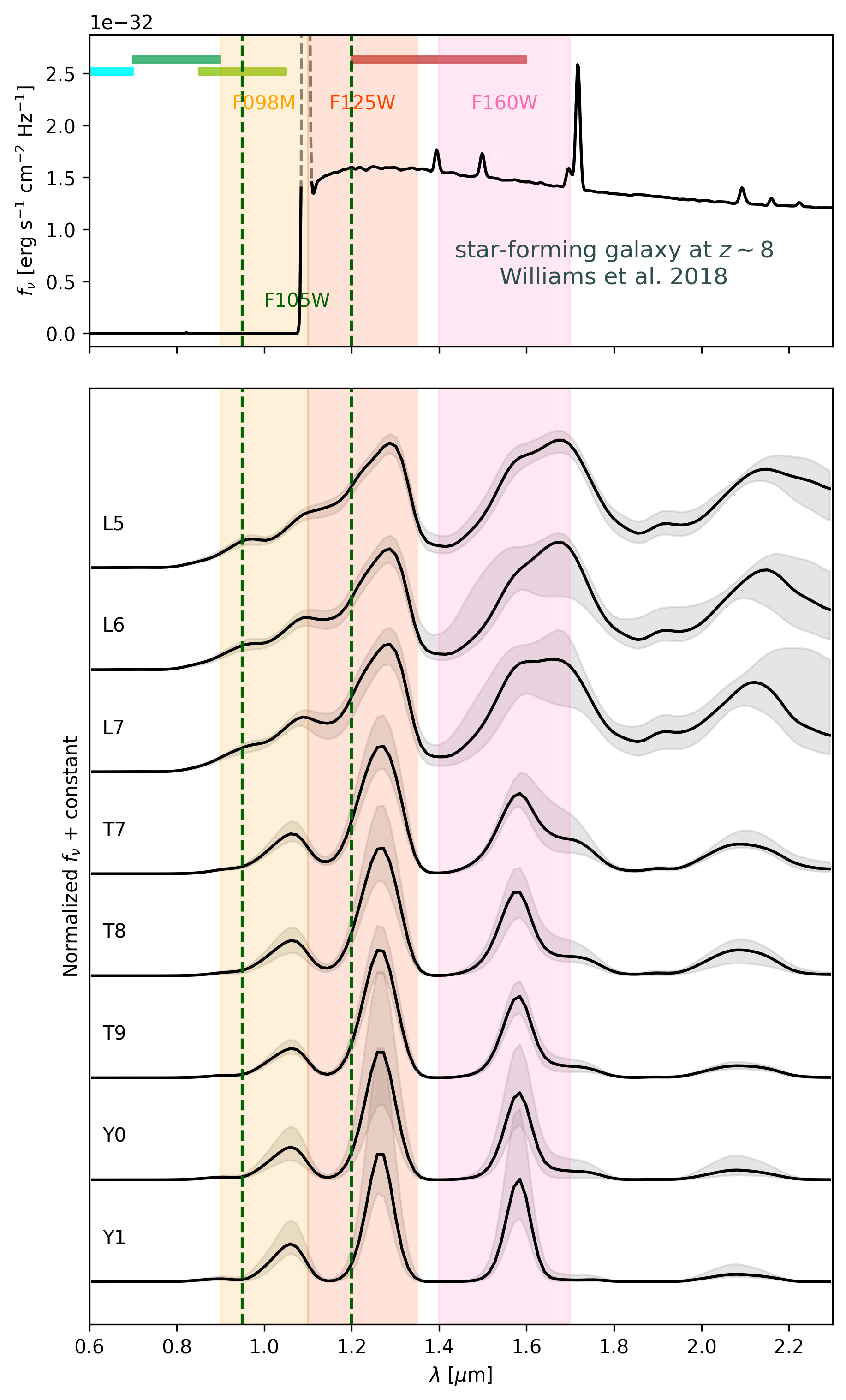}
    \end{minipage}

    \caption{Left panel: The color-color selection criteria for $z\sim8$ galaxies from \citet{borsani_2022}: F098M dropout (upper) and F105W dropout (lower). All gray data points are the simulated UCDs at $m_{F160W}\simeq26.5$ according to the characteristics of the BoRG-0409-5317 field. The colored data points are those that pass all selection criteria. F105W dropouts exhibit higher contamination than F098M dropouts, due to a less stringent criterion on the Y axis in the color--color space. Right panel: (Top) Example synthetic spectral energy distributions (SEDs) of a star-forming galaxy at $z \sim 8$ from \citet{Williams2018}. Shaded region and dashed lines represent the wavelength ranges of F098M, F125W, F160W, and F105W filters. The upper shaded regions correspond to additional filters: F606W, F814W, F850LP, and F140W. (Bottom) Median SEDs of the dominant UCD contaminant types that pass all $z\sim8_{F105W}$ galaxy selection criteria at $m_{F160W}\simeq26.5$.}
    \label{fig:color_098_105}
\end{figure*}

We simulate synthetic UCDs based on the depths of the BoRG-0409-5317 field, which includes both F098M and F105W observations with comparable $5\sigma$ limiting magnitudes, at 26.9 and 27.1 mag, respectively. Other filters used in the analysis are F606W, F814W, F125W, F140W, and F160W with $5\sigma$ limiting magnitudes of 26.35, 25.55, 27.02, 27.09, 26.75, respectively. In the left panels of Figure~\ref{fig:color_098_105}, the gray points represent simulated UCDs at $\mathrm{F160W} \simeq 26.5$ mag. The colored points indicate those that satisfy all selection criteria, including the photometric redshift constraint, signal-to-noise ratio requirements, and the $\chi^2$ comparison. 

From the figure, it is evident that a larger fraction of UCDs satisfy the F105W color selection compared to the F098M selection. Most UCDs have similar magnitudes in both F105W and F098M, indicating that the difference in contamination rates is due to the criteria involving the F098M/F105W$-$F125W color, where the threshold in one case is more permissive than the other.  

We further investigate the SEDs of the UCDs that satisfy all $z\sim8_{F105W}$ galaxy selection criteria in the right panels of Figure \ref{fig:color_098_105}. We find that the contaminants are predominantly ultra-cool dwarfs with spectral types late L and later ($T_\text{eff}\lesssim1600$K). These dwarfs show strong FeH molecular absorption bands around 1 \micron \citep{Geballe2002}, resulting in a similar integrated flux in the F098M and F105W filters. However, for $z\sim8$ galaxies (top panel), the F105W filter still includes part of the Lyman continuum. As a result, the F105W$-$F125W dropout criterion must be less stringent than the corresponding F098M$-$F125W dropout criterion, allowing substantially more UCD contaminants to enter the selection.

\section{UCD Contaminant Estimation for Other Space Telescopes}
\subsection{Predictions for the Roman Nancy Grace Deep Field Survey} \label{sec:Roman}
The \textit{Nancy Grace Roman Space Telescope} is a Hubble-class observatory with a field of view approximately 200 times larger than that of Hubble. It is equipped with a set of filters broadly similar to Hubble’s wide-band filters, with the addition of two bands centered at 1.84 and 2.13~$\mu$m. One of its science goals is to characterize the epoch of reionization. Given the large number of high-redshift sources it is expected to detect, it is essential to understand the properties and potential contamination from ultra-compact dwarfs (UCDs).

Here we predict the expected number of UCD contaminants in each field in each redshift range, using the planned deep-tier observations of the High-Latitude Wide-Area (HLWA) Survey\footnote{\url{https://roman-docs.stsci.edu/roman-community-defined-surveys/high-latitude-wide-area-survey}} as an example. The deep-tier observations will include two fields, centered at the COSMOS and XMM-LSS fields, each covering approximately $9.5~\mathrm{deg}^2$. 
The imaging will be performed using the F087, F106, F129, F146, F158, F184, and F213 filters with $5\sigma$ limiting magnitudes of 27.7, 27.7, 27.6, 28.3, 27.5, 27.0, 25.9.
These depths are approximately 0.5 mag deeper than the Hubble's NIR CANDELS survey.

To generate mock UCDs with FC-ENZO, we require a quantity analogous to \texttt{PHOTFLAM}, which converts the flux of a source into electrons per second and is used to determine the noise properties of the simulated sources. We calculate this \texttt{PHOTFLAM}-equivalent using the relation between SNR and magnitude in Equation \ref{eq: Flam_roman}. The SNR values are obtained from the Pandeia for Roman Exposure Time Calculator\footnote{\url{https://roman-docs.stsci.edu/simulation-tools/roman-wfi-exposure-time-calculator/pandeia-for-roman}}, adopting the survey exposure time per dither (294~s), together with the corresponding number of dithers and passes. 

We apply the following $z > 7$ galaxy selection criteria, constructed following the \citet{morishita_beacon_2024}:

\texttt{F087-dropouts} ($7.1 \lesssim z \lesssim 8.6$)
\begin{center}
$\mathrm{S/N}_{F106} > 4$ \\
$\mathrm{S/N}_{F087} < 2$ \\
$\mathrm{z}_\text{set} = 6.0 $ 
\end{center}

\texttt{F106-dropouts} ($8.6 \lesssim z \lesssim 10.5$)
\begin{center}
$\mathrm{S/N}_{F129} > 4$ \\
$\mathrm{S/N}_{F087, F106} < 2$ \\
$\mathrm{z}_\text{set} = 7.3 $ 
\end{center}

\texttt{F129-dropouts} ($10.5 \lesssim z \lesssim 12.8$)
\begin{center}
$\mathrm{S/N}_{F158} > 4$ \\
$\mathrm{S/N}_{F087, F106, F129} < 2$ \\
$\mathrm{z}_\text{set} = 8.9 $ 
\end{center}

\texttt{F158-dropouts} ($12.8 \lesssim z \lesssim 15.0$)
\begin{center}
$\mathrm{S/N}_{F184} > 4$ \\
$\mathrm{S/N}_{F087, F106, F129, F158} < 2$ \\
$\mathrm{z}_\text{set} = 10.9 $ 
\end{center}

\texttt{F184-dropouts} ($15.0 \lesssim z \lesssim 17.4$)
\begin{center}
$\mathrm{S/N}_{F213} > 4$ \\
$\mathrm{S/N}_{F087, F106, F129, F158, F184} < 2$ \\
$\mathrm{z}_\text{set} = 13.3$ 
\end{center}

Photometric redshifts are estimated using \texttt{EAZY} with the \citet{Hainline2024} template set, adopting a fitting range of $0.0 < z < 20$. We require that the best-fit redshift lies within the expected dropout redshift interval and that the integrated probability $P(z \ge z_\text{set})$ exceeds 0.8. The $\chi^2$ value of the best-fit galaxy template is then compared against the $\chi^2$ value of the best fit dwarf templates from \texttt{SpeX library \citep{Burgasser_spexprism}}. These selections essentially rely on photometric redshift constraints and non-detection in filters blueward of the Lyman break.

The results are shown in Figur~\ref{fig:roman_result}, where the expected contaminants are per $0.281~\mathrm{deg}^2$, corresponding to the field of view (FoV) of one pointing. For the F087 selection criteria, the contamination peaks at $F158\sim27$, which corresponds to the 5$\sigma$ detection limit of the band. The total contamination is estimated to be $20\pm4$ objects per field. In contrast, the F106 $z\sim9.5$ selection criteria yield a significantly lower contamination level of $0.51\pm0.18$ objects per field. The contaminations at other redshifts are essentially zero.  

\begin{figure}
    \centering
    \includegraphics[width=0.9\linewidth]{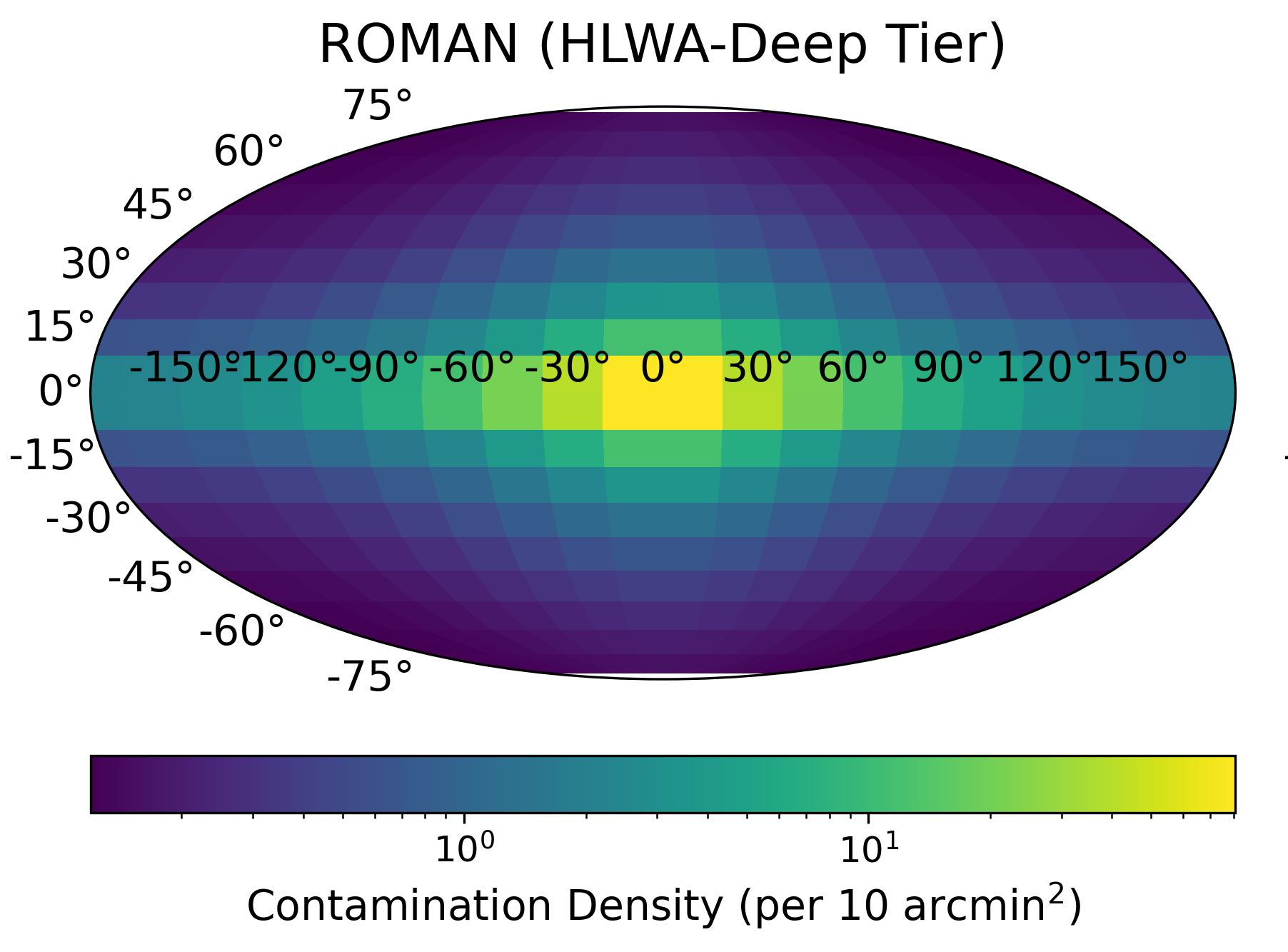}
    \caption{Predicted contamination map for the Nancy Grace Roman Space Telescope in units of per $10~\mathrm{arcmin}^2$ for $z\sim8$ galaxies using the characteristics of the HWLA survey and selection criteria presented in Section \ref{sec:Roman}.}
    \label{fig:ST_conta_map}
\end{figure}

We additionally present the predicted contamination map for $z\sim8$ dropouts in the unit of stars per 10 arcmin$^2$ with the characteristics of the HLWA survey, but move to different galactic coordinates in Figure~\ref{fig:ST_conta_map}. For the extragalactic field with $b>\ang{30}$, the contamination rate is generally smaller than 0.2 stars per 10 arcmin$^2$. 

\begin{figure}
    \centering
    \includegraphics[width=1.0\linewidth]{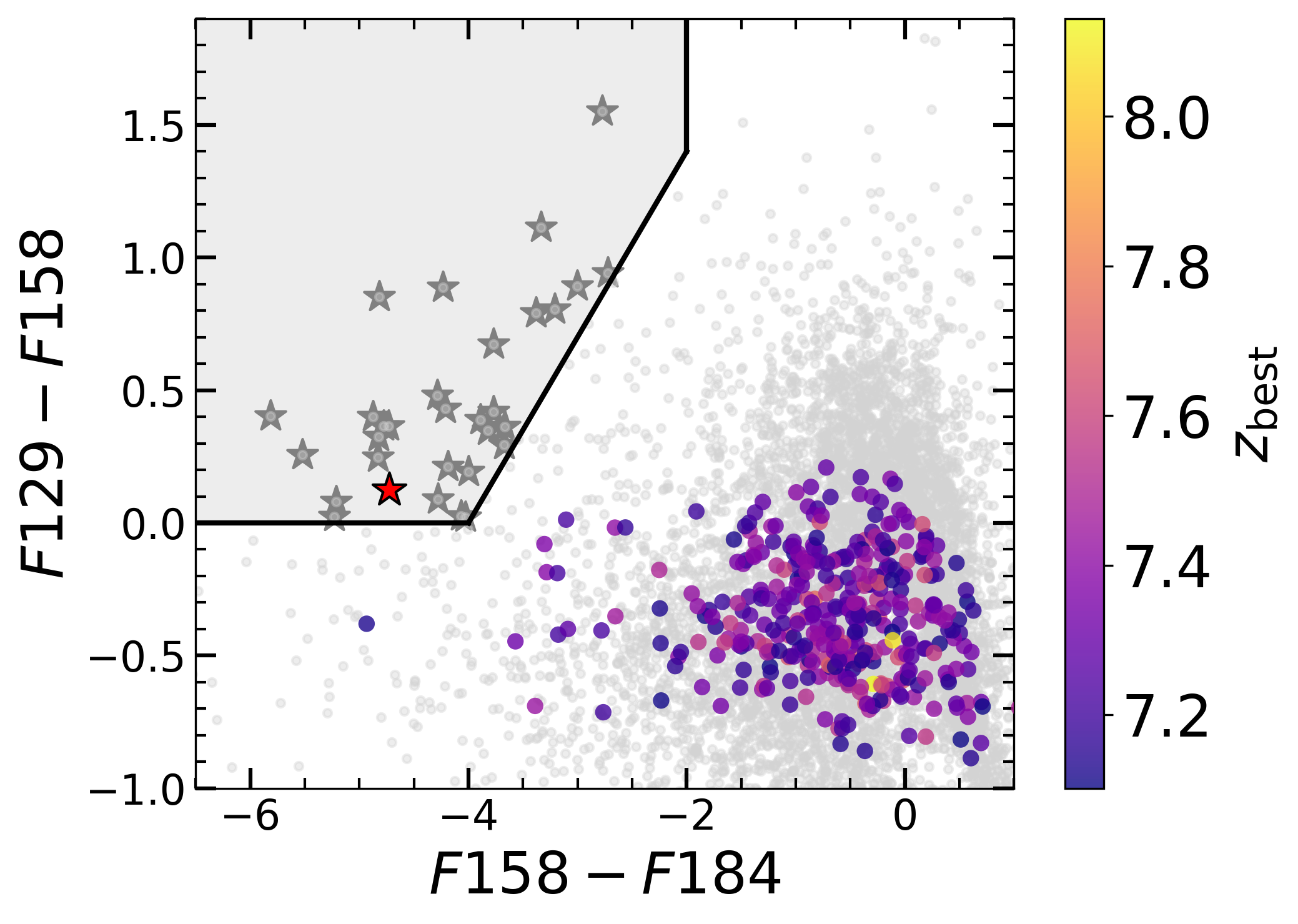}
    \caption{Stellar contamination removal using the criteria from \citet{Bagley2026}. All data points are UCDs simulated at F158=27.5 mag. The gray stars represent sources that fall within the stellar selection region defined by \citet{Bagley2026}. Colored data points are the remaining stellar contaminants in $z\sim8$ F087-dropout galaxy candidates.}
    \label{fig:bagley_result}
\end{figure}

\citet{Bagley2026} proposed two additional color-color selections with the Roman/WFI filters to separate stellar contaminants from galaxies: F062$-$F087 versus F087$-$F106 and F129$-$F158 versus F158$-$F184. The latter applies to the $z\sim8$ selection, as the former involves two filters that are part of the non-detection criteria. We apply the F129$-$F158 versus F158$-$F184 selection in Figure \ref{fig:bagley_result}, where objects falling within the shaded gray region are excluded as UCDs. All data points in the figure represent simulated UCDs at $F158=27.5$~mag. The colored data points indicate the subset that satisfies the F087-dropout criteria, comprising approximately $7.5\%$ of the simulated sample. We find that the criteria proposed by \citet{Bagley2026} can successfully identify a fraction of stellar contaminants. However, most contaminants remain even after this additional color--color screening, confirming that late-type UCDs exhibit colors similar to those of $z\sim8$ galaxies. 

\subsection{Predictions for the JWST Surveys} \label{sec:JWST}
High-redshift candidates selected from JWST imaging are generally considered less susceptible to contamination by UCDs because of JWST's mid infrared coverage \citep[e.g.,][]{Ryan_2016, Casey2023} and its superior spatial resolution, which enables improved separation between point-like and extended sources. In this section, we use \texttt{FC-ENZO} to recheck this using the observation characteristics from the Bias-free Extragalactic Analysis for Cosmic Origins with NIRCam (BEACON) survey \citet{morishita_beacon_2024}, a \textit{JWST} Cycle~2 program. Specifically, we used the BEACON field ID 0959+0200, which is in the vicinity of the \textit{COSMOS} field we previously checked for the \textit{HST} and \textit{Roman} observations in Figure \ref{fig:hst_result} and Figure \ref{fig:roman_result}. The field 0959+0200 includes imaging in eight filters: F090W, F115W, F150W, F200W, F277W, F356W, F410M, and F444W, with $5\sigma$ limiting magnitudes of 28.3, 28.3,28.5, 28.6, 28.9, 28.9, 28.2, 28.5. respectively. We artificially add a F070W filter with $5\sigma$ limiting magnitude of 28.6 mag, same as the depth of the J1235 field presented in \citet{Morishita_2024}.

\begin{figure}
    \centering
    \includegraphics[width=1.0\linewidth]{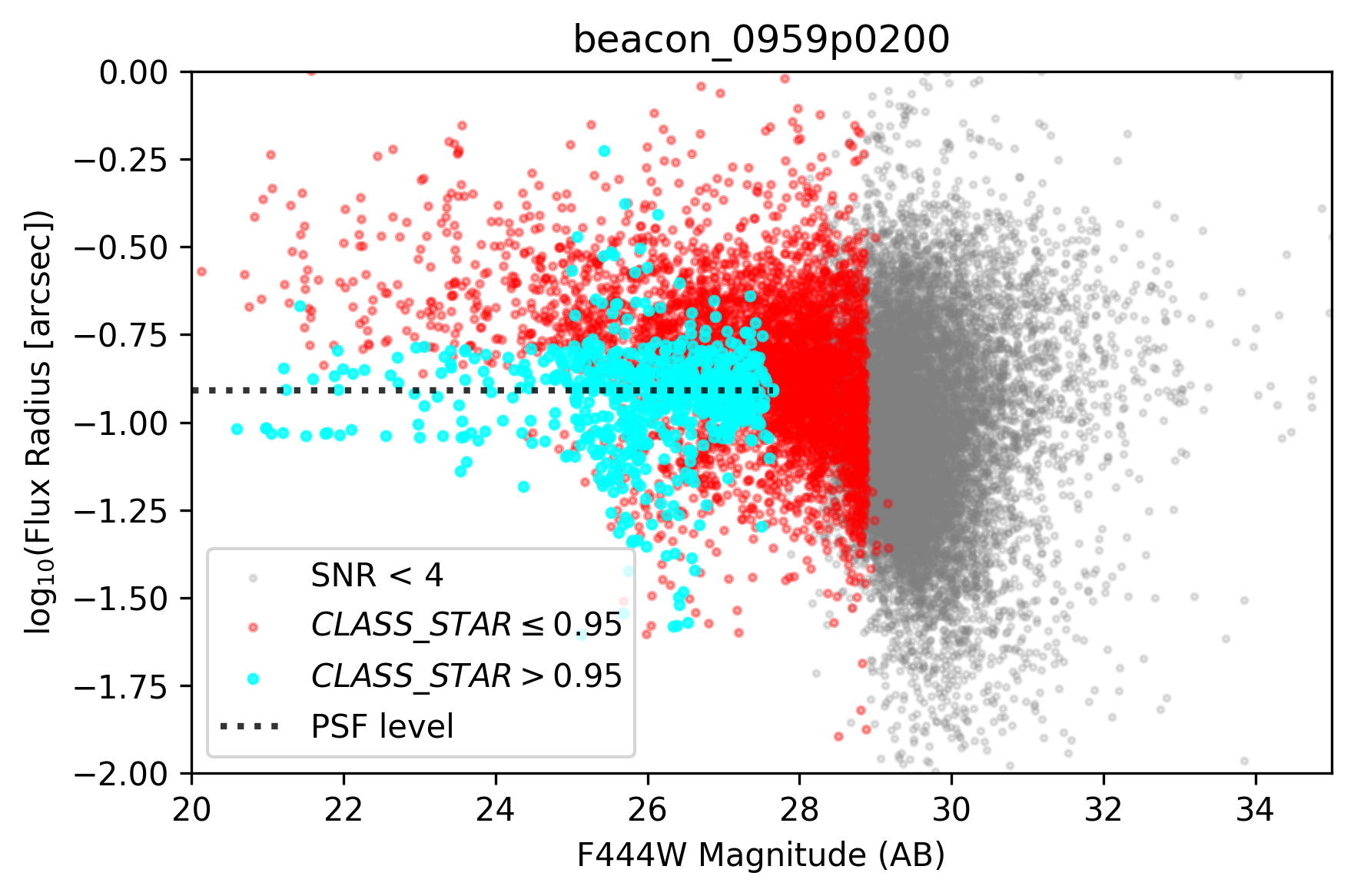}
    \caption{Distribution of sources classified using the \texttt{CLASS\_STAR} parameter derived from the F444W detection image. Gray points represent objects with signal-to-noise ratios lower than 4. Red points indicate extended sources, while cyan points indicate objects with \texttt{CLASS\_STAR} $> 0.95$, which are likely stellar sources. Morphological separation becomes increasingly difficult at magnitudes fainter than 27 mag.}
    \label{fig:beacon_class_star}
\end{figure}

We first determine the magnitude range over which point sources cannot be reliably distinguished from extended sources. To do so, we use the source catalog of the 0959+0200 field provided by \citet{morishita_beacon_2024}. Sources in this catalog were extracted using \texttt{SExtractor} \citep{Bertin_Arnouts_1996}, with the F444W image adopted as the detection band. In Figure~\ref{fig:beacon_class_star}, we plot the half-light radii against the F444W magnitudes for all detected sources with $\mathrm{S/N}_{\mathrm{F444W}} > 4$. Blue points indicate sources with \texttt{SExtractor}'s \texttt{CLASS\_STAR} parameter greater than 0.95, which typically identifies point-like objects. The stellarity parameter becomes unreliable for sources fainter than $\sim 27.5$ mag, while its reliability may already be reduced at brighter magnitudes.

We next simulate ultra-cool dwarfs (UCDs) over an F444W magnitude range of $24.0$ to $30.0$, capturing the whole range of candidates presented in \citet{morishita_beacon_2024},
using bins of $0.5$ mag. For \textit{JWST}, we instead calculate an equivalent of \texttt{PHOTFLAM} using the \texttt{PHOTMJSR} and \texttt{PIXAR\_SR} header parameters (Eq.~\ref{eq. photflam_jwst}). We fit the simulated photometry with the EA$z$Y code using the \citet{Hainline2024} template set to compute photometric redshifts and with the \texttt{SpeX} template sets while fixing the redshift at zero. We apply high-redshift galaxy selection criteria following the methods presented in \citet{Morishita_2024} and \citet{morishita_beacon_2024}, which can be summarized as follows:

\texttt{F070W-dropouts} ($5.0 \lesssim z \lesssim 7.3$)
\begin{center}
$\mathrm{S/N}_{115} > 8$ \\
$\mathrm{S/N}_{070} < 2.0$ \\
$\mathrm{z}_{set} = 4.5 $ 
\end{center}

\texttt{F090W-dropouts} ($7.3 \lesssim z \lesssim 9.7$)
\begin{center}
$\mathrm{S/N}_{150} > 4$ \\
$\mathrm{S/N}_{070, 090} < 2$ \\
$\mathrm{z}_{set} = 6.0$ 
\end{center}

\texttt{F115W-dropouts} ($9.7 \lesssim z \lesssim 13.0$)
\begin{center}
$\mathrm{S/N}_{200} > 4$ \\
$\mathrm{S/N}_{070, 090, 115} < 2$ \\
$\mathrm{z}_{set} = 8.0$ 
\end{center}

\texttt{F150W-dropouts} ($13.0 \lesssim z \lesssim 18.0$)
\begin{center}
$\mathrm{S/N}_{277} > 4$ \\
$\mathrm{S/N}_{070, 090, 115, 150} < 2$ \\
$\mathrm{z}_{set} = 10.0$ 
\end{center}


We also extend these criteria to the F200W, F277W, and F356W dropouts, corresponding to $18.0 \lesssim z \lesssim 24.5$, $24.5 \lesssim z \lesssim 31.0$, and $31.0 \lesssim z \lesssim 40.0$, respectively. As for the criteria listed above, the non-detection criteria include the dropout filter and all shorter-wavelength filters. The detection criterion of $S/N>4$ is applied to the F356W, F444W, and F444W bands, with $z_\text{set}=15$, 20, and 25, respectively.

We use the \texttt{SpeX} templates for the $\chi^2$ comparison. As mentioned earlier, several studies skip this $\chi^2$ comparison step when analyzing JWST-selected samples \citep[e.g.,][]{Whitler2025,Weibel2026,Leethochawalit2026}. Although \citet{Finkelstein2024} uses the Sonora-Diamondback models \citep{Diamondback2024} as templates for brown dwarf fitting, we opt to use the \texttt{SpeX} templates to remain consistent with the other criteria presented above. Moreover, the Sonora-Diamondback models are derived from the same parent code as the Sonora Elf Owl models used to generate the potential contaminants in our analysis. Using a similar set of spectra for both generating the contaminants and fitting the candidates could therefore introduce a bias in the results. Because the \texttt{SpeX} templates extend only to $2.5\,\micron$, we calculate the $\chi^2$ for the best-fit \citet{Hainline2024} template using only filters with effective wavelengths shorter than $2.5\,\micron$ in the $\chi^2$ comparison. Consequently, we cannot perform the $\chi^2$ comparison for the F277W and F356W dropout samples.

After applying these criteria, we estimate the expected UCD contamination across different redshift intervals and present the results in Figure~\ref{fig:compare_surveys}. In the last panel, the dashed lines indicate the contaminant rates without the $\chi^2$ comparison, while the solid lines show the corresponding rates when the $\chi^2$ comparison is included. Two dropout selections, F070W and F150W, yield zero UCD contamination rates both with and without the $\chi^2$ comparison, while the contamination rate for F200W is below $2.5\times10^{-3}$ contaminants per field without the $\chi^2$ comparison. We exclude these three dropout selections from the plot. Without the $\chi^2$ comparison, the expected contamination rates are 1.65, 0.01, 0.10, and 1.3 contaminants per field for the F090W, F115W, F277W, and F356W dropout selections, respectively. When the $\chi^2$ comparison is included, the contamination rates for the F090W and F115W dropout selections decrease to 0.14 and $6.5\times10^{-3}$ contaminants per field, respectively. 

The contamination fractions of the F090W dropouts remain broadly consistent with those found in the HST results, although the contaminants shift toward fainter magnitudes near the detection limits of the JWST observations. A more surprising result is the non-negligible contamination fraction of the F356W dropouts ($z \gtrsim 30$). These contaminants are primarily associated with Y dwarfs, whose spectra are dominated by an emission bump at $\sim 4\,\mu$m caused by the combined effects of $\text{H}_2\text{O}$ and $\text{CH}_4$ absorption features, which can mimic the dropout feature expected for $z>30$ galaxies. Our result is consistent with the identification of two Y dwarfs that were subsequently followed up after initially being considered possible $z\sim20$--30 galaxy candidates with F444W $\sim 27$ mag, as reported by \citet{bradac_2026}. While the first generation of stars is expected to form at $z\sim30$ \citep{Klessen2023}, the high prevalence of contaminants found here suggests that any F356W dropout candidates identified in JWST/NIRCam data are likely to be UCD contaminants. Deep JWST/MIRI observations will therefore be necessary to identify high-redshift galaxy candidates.

Adding medium bands to JWST/NIRCam observations can help exclude contamination from mid-redshift Balmer-break galaxies from the high-redshift sample \citep{Eisenstein2025}. We now test whether these medium bands can also help with UCD contaminations. We utilize deep medium band characteristics from the GOODS-N field of the JWST Advanced Deep Extragalactic Survey \citep[JADES: ][]{Eugenio2025}, specifically adding F128M, F210M, and F355M filters with depths of 28.77, 28.50, and 29.45~mag, respectively, to the original characteristics of wide band filters in the BEACON field.

We find that adding medium bands can significantly reduce UCD contamination in F090W dropout candidates. Without the $\chi^2$ comparison, the contamination rate decreases by approximately a factor of two, while with the $\chi^2$ comparison, it decreases by approximately a factor of three. This is because the medium bands both help distinguish UCDs from high-redshift galaxy in the SED $\chi^2$ comparison and shift the best-fit redshifts obtained with the galaxy templates toward $z\sim1$, such that the UCDs no longer satisfy the $p(z)$ criteria. However, we find no significant change in the contamination rate of F356W dropouts, because the additional medium bands fall within the non-detection wavelength ranges, while the Y-dwarf contaminants have very little $f_\nu$ flux at wavelengths shorter than $\sim3.5\,\mu$m. We confirm this using the depths of both the broad- and medium-band filters in the JADES survey and obtain similar results, although the peak magnitude of the contaminants shifts to fainter magnitudes because the JADES survey is, overall, deeper than the BEACON survey.

\subsection{Predictions for the \textit{Euclid} Deep Field Survey}

The first \textit{Euclid} data release \citep{Euclid_Collaboration_2026} covers a total area of $63.1~\mathrm{deg}^{2}$ and contains approximately 26 million detected sources. The current data set covers all the areas of the three deep fields but at the wide-survey depths. The full data set is expected to reach the $5\sigma$ depth at 2~mag deeper than the current data set, i.e., reaching 28.2, 26.3, 26.5, and 26.4~mag for the \textit{VIS}, \textit{Y}, \textit{J}, and \textit{H} filters, respectively.

In this section, we predict the number of UCD contaminants in the \textit{Euclid} Deep Fields. To estimate the expected contamination in these observations, we utilize the selection criteria as follows:

\texttt{Y-dropouts} ($7.7 \lesssim z \lesssim 10.1$)
\begin{center}
$\mathrm{S/N}_\text{H} > 4$ \\
$\mathrm{S/N}_\text{VIS} < 2$ \\
$\mathrm{z}_{set} = 6.7$ 
\end{center}

\texttt{J-dropouts} ($10.1 \lesssim z \lesssim 13.11$)
\begin{center}
$\mathrm{S/N}_\text{H} > 4$ \\
$\mathrm{S/N}_\text{VIS, Y} < 2$ \\
$\mathrm{z}_{set} = 8.4$ 
\end{center}

\texttt{H-dropouts} ($13.11 \lesssim z \lesssim 15.9$)
\begin{center}
$\mathrm{S/N}_\text{H} > 4$ \\
$\mathrm{S/N}_\text{VIS, Y, J} < 2$ \\
$\mathrm{z}_{set} = 11.3$ 
\end{center}
The standard $\chi^2$ comparison is applied as usual.

We show in Figure~\ref{fig:euclid_result} the expected contamination within an FOV ($0.53~\mathrm{deg}^2$) for the \textit{Y}-dropout sample ($7.7 \lesssim z \lesssim 10.1$) at the location of the EDF-North (EDFN) field. The total expected contamination is $90.4 \pm 16.5$ objects. Similar to our previous findings, the peak of the contamination distribution consistently coincides with the $5\sigma$ depth of the detection band. Although the differences are not statistically significant at the $2\sigma$ level, EDFN exhibits the highest predicted contamination, followed by EDF-Fornax (EDFF) and EDF-South (EDFS).

\subsection{Comparison across Surveys and Interpretation toward Luminosity Function Determination}

\begin{figure}
    \centering
    \includegraphics[width=1.0\linewidth]{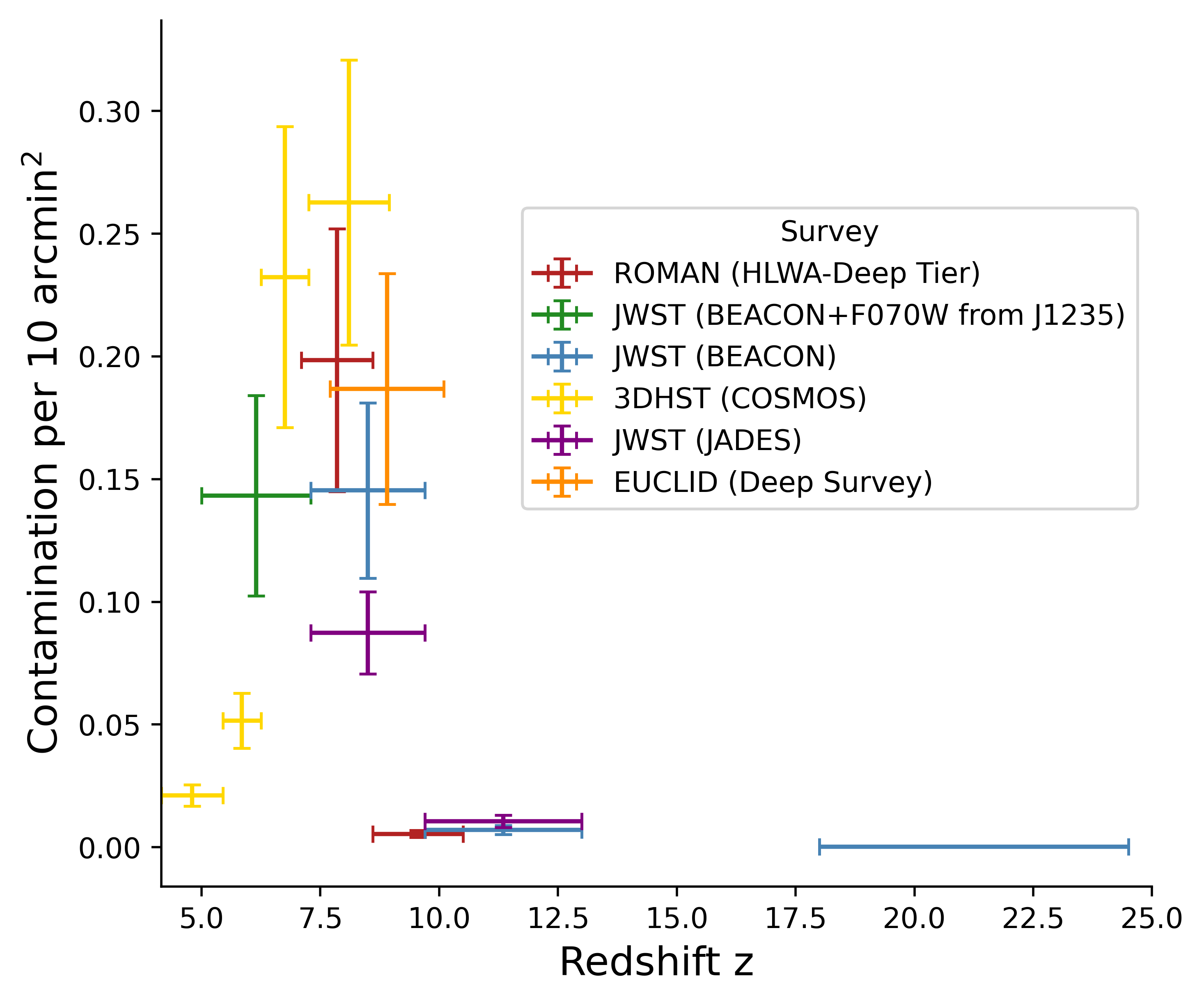}
    \caption{Predicted contamination rates normalized to an area of $10\,\mathrm{arcmin}^{2}$ for different surveys and redshift selections. The horizontal error bars indicate the redshift range associated with the selection criteria given in the text. The fields are the following: red -- the Roman field ROMAN-1000+0211; blue -- JWST-BEACON-0959+0200; green -- JWST-BEACON-0959+0200 with added F070W filter; yellow -- HST-COSMOS-1000+0224; purple -- JWST-JADES GOODS-N; and orange -- Euclid -- EDFN Deep Field. All fields are located in the vicinity of the COSMOS region except JADES GOODS-N and EDFN.}
    \label{fig:compare_surveys}
\end{figure}

\begin{figure}
    \centering
    \includegraphics[width=1.0\linewidth]{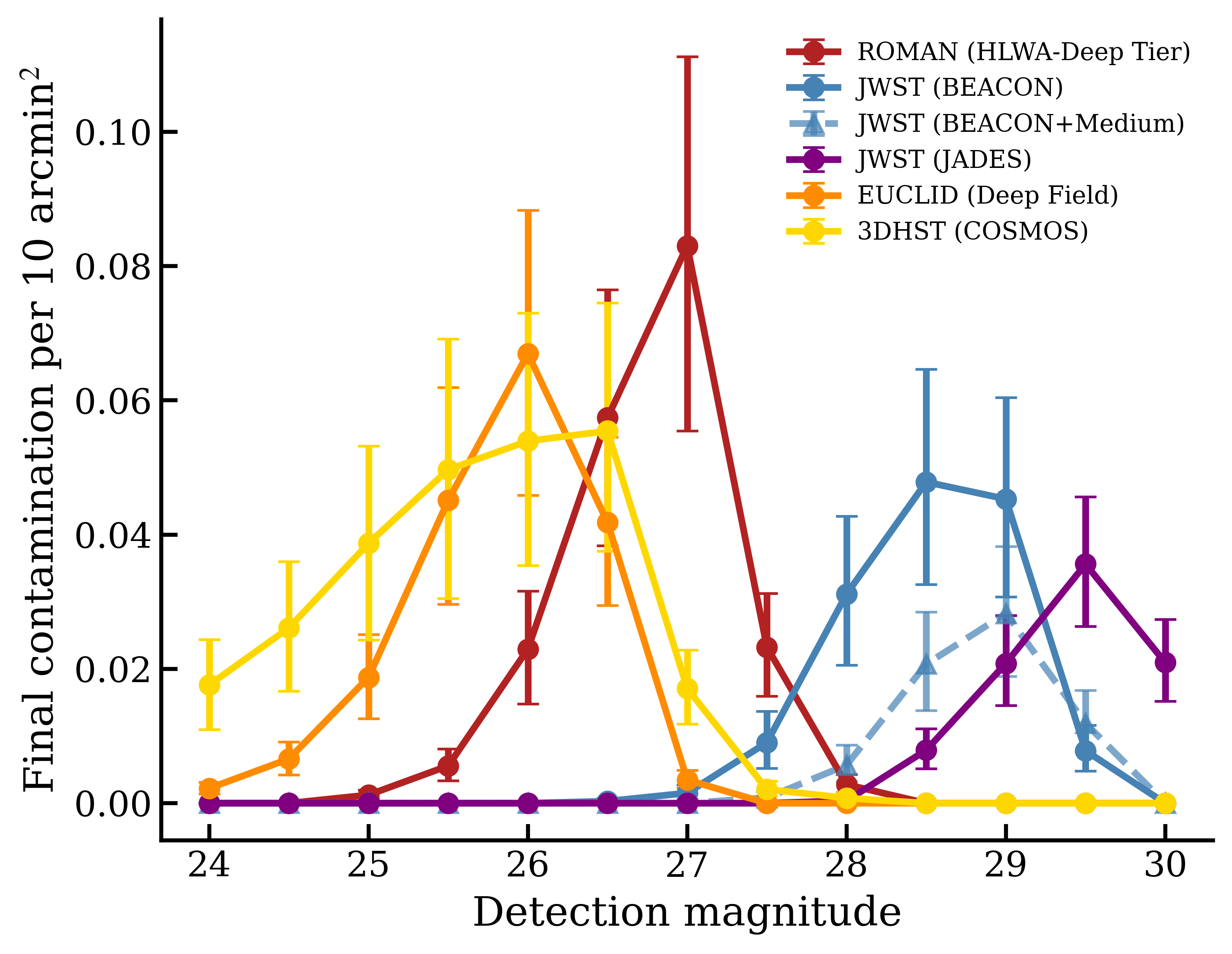}
    \caption{Contamination rates for $z \sim 8$ as a function of apparent magnitude. The color scheme and the survey descriptions are the same as in Figure~\ref{fig:compare_surveys}. The blue dashed line represents the JWST-BEACON-0959+0200 survey, with three additional medium bands from JADES.}
    \label{fig:mag_sep}
\end{figure}

In Figure~\ref{fig:compare_surveys}, we plot all contamination rates we explored in this paper normalized to the same survey area of $10\,\mathrm{arcmin}^{2}$ as a function of redshift. This plot further emphasizes that $z\sim8$ is the redshift range most prone to contamination from UCDs, regardless of the telescopes. We also plot the contamination rates at $z\sim8$ as a function of detection magnitude (F160W, F158, F444W, and \textit{H} bands) in Figure \ref{fig:mag_sep}. The fields are the \textit{HST}, \textit{JWST}, and \textit{Roman} observations in the COSMOS area, alongside \textit{Euclid} observations in the EDFN field. Although the overall contamination levels at a given redshift are similar across different telescopes (Figure \ref{fig:compare_surveys}), Figure \ref{fig:mag_sep} shows that the contaminants arise from different magnitude ranges, typically near the detection limits of the respective surveys.

\begin{figure}
    \centering
    \includegraphics[width=1.0\linewidth]{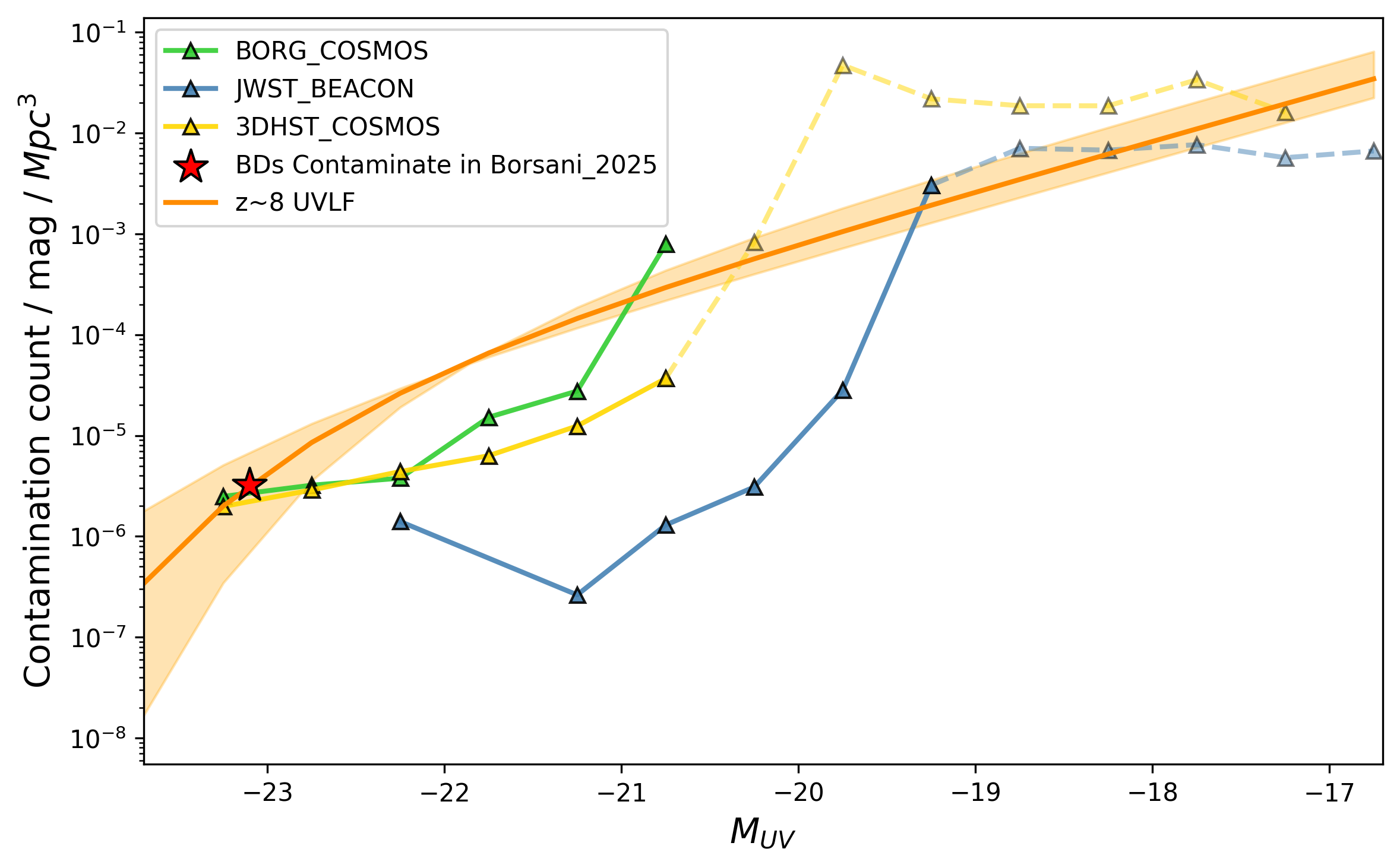}
    \caption{$z\sim8$ UCD contamination counts translated to units of per mag per Mpc$^3$, using the surveys explored in this work, but excludes results from \textit{Roman} due to the lack of available completeness estimate. We also compare the contamination with the best-fit $z \sim 8$ galaxy luminosity function from \citet{leethochawalit_uv_2023}, shown as the orange curves. The red star at $M_{\mathrm{UV}} = -23.10$ corresponds to a T2 dwarf identified as a contaminant in spectroscopic follow-up observations by \citet{borsani_2025}. The dashed lines at the faint end of each plot indicate the regime where the completeness drops by $30\%$, typically discarded during luminosity function calculation.}
    \label{fig:Mpc3_con}
\end{figure}

We further translate these contamination rates into luminosity functions, i.e., the number counts per magnitude per Mpc$^3$ arising solely from UCD interlopers. The results are shown in Figure~\ref{fig:Mpc3_con} for the HST BORG-0955+4528 field, the HST COSMOS field, and the JWST BEACON-0959+0200 field. We use the completeness function for the BoRG field from \citet{leethochawalit_uv_2023}. The completeness functions for the other fields are calculated using the completeness code from \citet{Leethochawalit2022}, together with the selection criteria from \citet{bouwens_2015} and \citet{Morishita_2024} for the HST COSMOS and JWST BEACON fields, respectively. For comparison, we also show the $z \sim 8$ galaxy luminosity function from \citet{leethochawalit_uv_2023} as the solid orange line.

Figure~\ref{fig:Mpc3_con} shows a potentially concerning result at the bright end, where the contaminant number densities from relatively shallow surveys are of the same order of magnitude as the expected galaxy number density, assuming our fiducial UCD contamination model. At fainter magnitudes, the contamination rates in HST surveys are approximately $0.5$--$1$ dex lower than the galaxy number density. The results for JWST are much more promising, generally at a few dex below the galaxy number density. Although the contamination rates rise again near the magnitude limits in all surveys, this is less concerning because these regimes correspond to where the completeness is low and are typically excluded from luminosity function calculations. Our results, therefore, highlight the importance of deep surveys in constraining the bright end of the $z\sim8$ luminosity function.

The choice of field strategy also has implications for UCD contamination. Unlike cosmic variance, which can be mitigated by combining observations from multiple independent sightlines \citep{Trenti2008}, UCD contamination depends strongly on the Galactic position of each field (see Figure \ref{fig:ST_conta_map}). In particular, fields closer to the Galactic plane and those at $|l|\sim0^\circ$ have higher contamination. Thus, a random distribution of extragalactic pointings is not necessarily optimal for minimizing UCD contamination, while a field selected to lie in a region of low predicted UCD density can have lower contamination than a random set of pointings.


\section{CONCLUSIONS}
In this work, we present \texttt{FC-ENZO}, a code developed to estimate the magnitude-dependent contamination from ultra-cool dwarf stars (UCDs) that can mimic high-redshift galaxy candidates. The code can be used to design observations or identify fields with potentially high contamination rates prior to spectroscopic follow-up observations. \texttt{FC-ENZO} operates at the mock-catalog and survey-characteristic level by combining UCD number density models, synthetic spectral libraries, and galaxy selection methods. 

We test the robustness of the code against different model assumptions. We also apply galaxy selection criteria across a range of redshifts and to representative surveys from HST, JWST, EUCLID, and the Nancy Grace Roman telescopes. Our main findings are as follows.

\begin{enumerate}
    \renewcommand{\labelenumi}{(\roman{enumi})}
    \item Absolute contamination rates are sensitive to assumptions in the model, such as the adopted stellar evolution models, metallicities, and stellar distribution models, leading to variations of up to several tens of a factor in the predicted contamination rates. However, the former sources of uncertainty are less concerning. First, the adopted fiducial stellar evolution model, the \textit{Elf Owl} evolutionary model, contains much larger number of spectral templates, approximately 6 times more than \textit{Bobcat}, the comparison model used in our tests. This broader coverage results in a higher representation of T- and Y-type dwarfs, which are the dominant contaminants in high-redshift galaxy selections. In addition, \textit{Elf Owl} does not assume chemical equilibrium. Second, the metallicities of the contaminating UCDs are likely to be solar or subsolar, reducing the sensitivity of the results to the assumed metallicity distribution. In contrast, the contamination rates remain strongly sensitive to the adopted number density model, with variations of up to several factors, highlighting the need for better constraints on the Galactic distribution of UCDs. Our fiducial model, adopted from \citet{aganze_2022b}, is among the most up-to-date and incorporates a more complex Galactic structure, including thin-disk, thick-disk, and halo components that capture the distribution of T to early-Y type UCDs. 
    
    \item When tested on the HST BoRG fields with JWST spectroscopic follow-up observations, our code identifies the field confirmed to contain UCD contaminants as the field most likely to be contaminated. This demonstrates that the code can effectively rank fields and help identify the most suitable candidates for spectroscopic follow-up observations.
    
    \item The redshift range with the highest level of contamination is $z \sim 8$, regardless of the telescopes. For the $z \sim 8$ selection criteria with the HST/WFC3 instrument, the F105W dropout selection yields a higher contamination rate compared to the F098W dropout. This is because the NIR color selection in the F105W dropout is less stringent on the Lyman-break color, which allows more UCDs contamination. Among all UCD spectral types, late L to early-Y dwarfs are the primary contaminants, as their spectral energy distributions closely resemble those of galaxies in the redshift range $7 \lesssim z \lesssim 8.5$.

     \item For the same survey area, all space telescopes with broad-band observations exhibit comparable numbers of contaminants at a given redshift. However, the magnitude range over which the contaminants appear varies and typically peaks near the limiting magnitude of the survey. As a result, shallow surveys tend to have a higher fraction of contaminants at the bright end. Adding medium-band filters can further reduce the UCD contamination rate at $z\sim8$ by a factor of two to three. When applying the completeness correction to express the contamination in the form of a luminosity function, we find that the contamination from UCDs becomes comparable to the $z \sim 8$ galaxy luminosity function at the bright end for \textit{HST} surveys. In contrast, deeper surveys conducted with \textit{JWST} show significantly improved outcomes.

    \item At $z\gtrsim30$, F356W dropout selections can suffer substantial contamination from Y dwarfs, whose $\sim4,\micron$ spectral features can mimic the expected dropout signature of extremely high-redshift galaxies. This suggests that F356W dropout candidates require additional observations at longer wavelengths, such as with \textit{JWST}/MIRI, to better distinguish high-redshift galaxies from UCD contaminants.
\end{enumerate}

\begin{acknowledgments}
We are thankful to the reviewers and anonymous users of our code who provided valuable suggestions and feedback. This research made use of data obtained from the Mikulski Archive for Space Telescopes (MAST). This work is supported by the Fundamental Fund of Thailand Science Research and Innovation (TSRI) 1260 through the National Astronomical Research Institute of Thailand (Public Organization) (FFB690078/0269).

\end{acknowledgments}

\appendix
\section{Noise Modeling}
\label{apd:noise_flux_new_mag}

Since FC-ENZO is designed to operate at the survey-characteristics level, providing the flexibility to both aid observation design and statistically evaluate contamination after observations are obtained, we adopt a theoretical Poisson noise model when generating the UCD mock catalogs. The main parameters used are \texttt{PHOTPLAM}, \texttt{PHOTFLAM}, \texttt{EXPTIME}, and the limiting magnitudes (depths). Here, we test the performance of this noise model against real observations.

We calculate the noise perturbation starting from the Poisson noise expression:

\begin{equation}
    \mathrm{Poisson\ Noise}_{e} = \sqrt{N_{e}} = \sqrt{S_{e} + B_{e}},
\end{equation}

where $S_{e}$ is the source flux in units of electrons. $B_{e}$ is the background variance contribution in units of electrons, which includes the sky background, dark current, and read noise. This term is equivalent to the square of the $1\sigma$ limiting flux in the band of interest, i.e., $B_{e} = \sigma_{1\sigma}^{2}$.

We can convert this equation to use the flux density per unit frequency $f_{\nu}$ in cgs units ($\mathrm{erg\ s^{-1}\ cm^{-2}\ Hz^{-1}}$) as

\begin{equation}
    \mathrm{Poisson\ Noise}_{e} =
    \sqrt{
        \frac{
            c \cdot f_{\nu} \cdot \texttt{EXPTIME}
        }{
            \texttt{PHOTPLAM}^{2} \cdot \texttt{PHOTFLAM} 
        }
        +
        \left(
        \frac{
            c \cdot f_{1\,\sigma,\nu} \cdot \texttt{EXPTIME}
        }{
            \texttt{PHOTPLAM}^{2} \cdot \texttt{PHOTFLAM} 
        }
        \right)^{2}
    }.
    \label{eq: Flam_roman}
\end{equation}

\texttt{PHOTPLAM}, \texttt{PHOTFLAM} and \texttt{EXPTIME} are common parameters in the header of HST images. Specifically, \texttt{PHOTFLAM} is the flux ($f_\lambda$) that would produce 1 electron per second, with a unit of  $\mathrm{(erg\ cm^{-2}\ s^{-1}\ \text{\AA}^{-1}})(e/s)^{-1}$. \texttt{PHOTPLAM} is the pivot wavelength in angstrom. \texttt{EXPTIME} is the total exposure time in seconds. Finally, the speed of light $c$ is expressed in $\text{\AA}\ \mathrm{s^{-1}}$.

We can write this quantity back to the cgs unit:
\begin{equation}
    \mathrm{Poisson\ Noise}_{\nu, \mathrm{cgs}} =
    \sqrt{
        \frac{f_\nu}{K}+f_{1\,\sigma,\nu}^2
    }.
\end{equation}
where K is $\frac{c \cdot \texttt{EXPTIME}}{\texttt{PHOTPLAM}^{2}\cdot\texttt{PHOTFLAM}}$.

\begin{figure}
    \centering
    \includegraphics[width=0.5\linewidth]{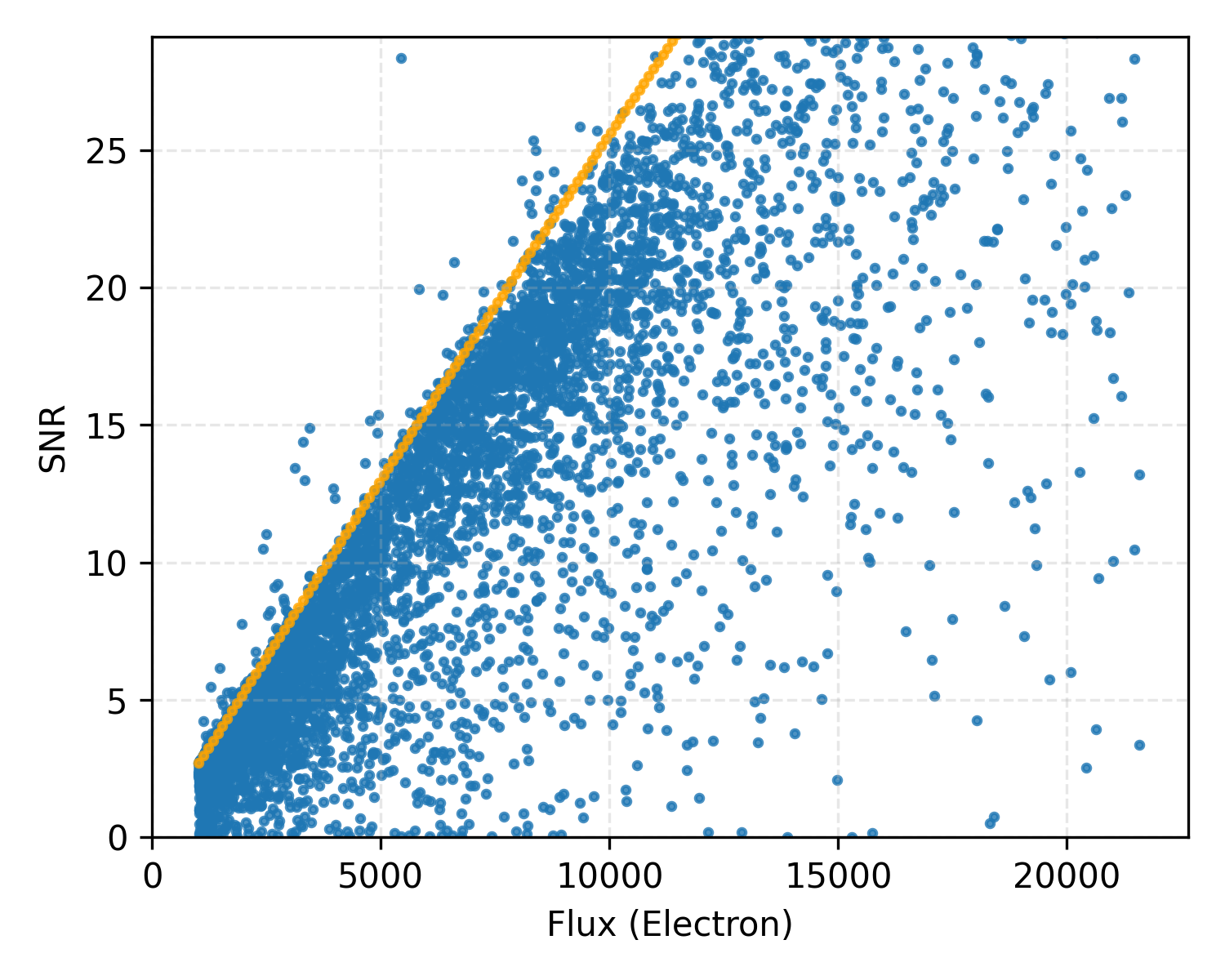}
    \caption{Relation between signal-to-noise ratio (SNR) and HST F160W flux from \texttt{BORG\_0132+3035}. The data points represent the observed SNR calculated from total flux, while the orange line shows the values derived from our noise calculation.}
    \label{fig:0132+3035_F160W_noise_cal}
\end{figure}

We validate this noise model using the catalog from the 
\texttt{BORG\_0132+3035} dataset and show the relation between S/N and flux in Figure~\ref{fig:0132+3035_F160W_noise_cal}. The fluxes and S/N values are from the total flux calculation in \citet{Morishita_2020}. The calculated S/N traces the upper envelope of the observed noise distribution, as expected for an idealized noise model.

\begin{figure}
    \centering
    \includegraphics[width=0.5\linewidth]{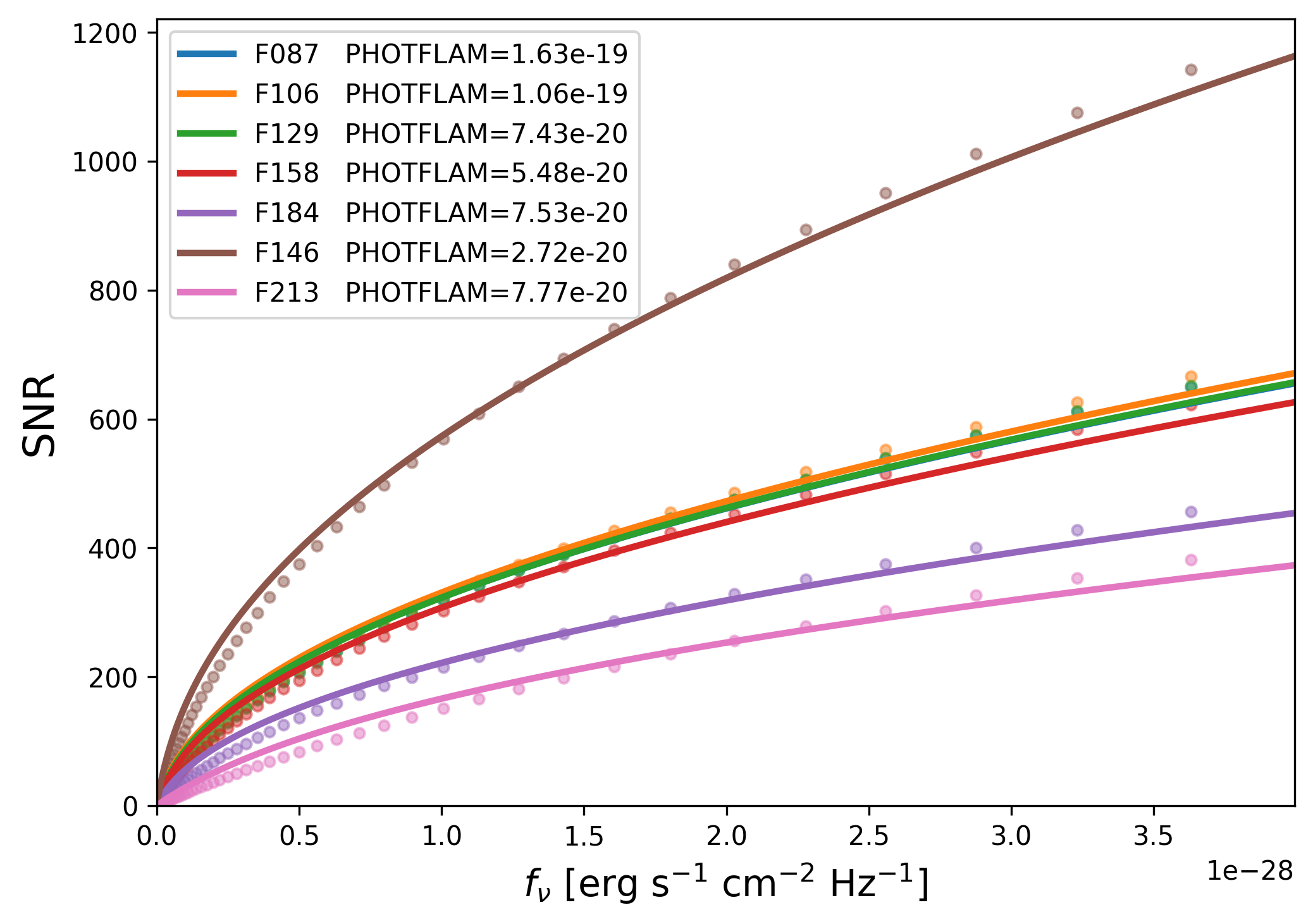}
    \caption{
    Fitting of the \texttt{PHOTFLAM} calibration parameter based on the \texttt{Pandeia ETC}. The data points were obtained from the Pandeia simulations, while the solid line shows the best-fit model. For the ETC, we used the exposure time corresponding to the Roman High-Latitude Wide-Area Survey (HLWAS) deep-tier broadband survey.
    }
    \label{fig:flam_roman}
\end{figure}

For the \texttt{Roman Space Telescope}, the S/N values were calculated using the \texttt{Pandeia} Exposure Time Calculator (ETC) \citep{Pandeia} and are shown as data points in Figure~\ref{fig:flam_roman}. We adopt the survey characteristics, including the MULTIACCUM readout pattern with an exposure time of 294~s per dither (MA table \texttt{IM\_294\_16}), with typically 3 dithers and 5 passes. Using these calculations, we determine the best-fit $K$ parameter and solve for the corresponding \texttt{PHOTFLAM} value. The best-fit SNR models are shown as a solid line in the figure. The derived \texttt{PHOTFLAM} values are listed in the legend of the figure.


For \texttt{JWST}, calibrated science images are typically provided in units of MJy/sr, unlike \texttt{HST}. Therefore, the \texttt{PHOTFLAM} parameter can be written as

\begin{equation}
    \texttt{PHOTFLAM}
    =
    \frac{
        \texttt{PHOTMJSR}
        \cdot
        \texttt{PIXAR\_SR}
        \cdot
        c
        \cdot
        10^{-17}
    }{
        \texttt{PHOTPLAM}^{2}
    }.
    \label{eq. photflam_jwst}
\end{equation}

where \texttt{PHOTMJSR} represents the flux density in MJy/sr that produces one count per second, while \texttt{PIXAR\_SR} is the nominal pixel area in steradians. Both parameters can be obtained directly from the \texttt{JWST} science image headers. 

\section{Limitations of the UCD Contamination Model}
\label{apd:limitation}
In Section \ref{sec:contamrate_and_modelparam}, we explored the sensitivity of the predicted UCD contamination to variations in the parameters of our fiducial model. Here, we consider additional limitations arising from effects that are not included in the model. In particular, we examine the potential effects of substructure in the UCD spatial distribution, including stellar streams, spiral arms, and north--south asymmetries, as well as stellar multiplicity.

The Milky Way UCD distributions adopted here assume a smooth spatial density distribution and therefore do not account for substructures, such as stellar streams and spiral arms, that can produce localized overdensities. As a result, our model may underestimate the UCD contamination rate in specific fields. For example, an overdensity of M dwarfs has been identified at the location of the Sagittarius stream in one of the BoRG fields \citep{holwerda2014}.

For stellar streams, we find that they are generally too distant to be a significant concern, particularly for surveys with detection limits of $m\lesssim27$ mag. At $\text{F160W}=27$ mag, M9 and cooler UCDs are located at distances of less than 15 kpc. In contrast, several massive stellar streams, including Sagittarius, Orphan--Chenab, Cetus--Palca, and Kwando, are located at distances of $\sim15$--100 kpc \citep{hernitschek_2017,Koposov2023,Thomas,galstreams}. For deeper surveys reaching $m\sim30$ mag, however, late M and L dwarfs can be located at distances overlapping with those of these stellar streams. Although most of the $z\sim8$ contaminants are expected to be late L and later UCDs (Section \ref{sec:HSTlesson}), $\sim10\%$ of the contaminants can still originate from late M and early L dwarfs. Therefore, overdensities of M and L dwarfs associated with stellar streams could provide an additional contribution to the $z\sim8$ contamination in deep surveys. We do not expect a significant contribution from stellar streams for $z\sim30$ F356W dropouts, since their contaminants are exclusively T and Y dwarfs, which remain at distances of less than 10 kpc even for sources as faint as F160W$=30$ mag.

Spiral arms are another form of Galactic substructure not included in our model. The Sun lies within the Local (Orion) Arm, while the two nearest major spiral arms, the Sagittarius--Carina and Perseus arms, have minimum distances of approximately 2~kpc from the Sun \citep[e.g.,][]{Reid2019}. An overdensity of $\sim1$~Gyr-old stars associated with the Local Arm has also been identified in Gaia data, particularly at Galactic longitudes of $l=90^\circ$--$110^\circ$, with an overdensity contrast reaching $\sim50\%$ \citep{Miyachi2019}. However, this overdensity is located at a distance of approximately 1~kpc \citep{Miyachi2019,Poggio2021}, while the vertical extent of these spiral-arm overdensities is less than $\sim0.1$~kpc \citep{Poggio2026}. At a distance of 1~kpc, this corresponds to a vertical angular extent of only $\sim6^\circ$, indicating that the overdensity is largely confined to low Galactic latitudes. Therefore, the contribution of spiral-arm overdensities to UCD contamination in extragalactic fields is expected to be small.

Another potential limitation is a north--south asymmetry in the spatial distribution of UCDs arising from our assumption of a fixed Galactic mid-plane. We adopt $Z_{\odot}=27$~pc \citep{Gillessen_2009,Chen_2001} for the Solar position relative to the Galactic mid-plane, a value that is typically fixed in Galactic structure models. However, the effective mid-plane traced by UCD populations may vary with Galactic position and may not be adequately described by a single global value. Observational evidence for such north--south differences has been reported for stellar populations. For example, \citet{Pirzkal2009} found different numbers of M dwarfs toward the PEARS-N and PEARS-S fields, while \citet{Iyisan2025} found slightly larger scale heights for evolved stars in the southern hemisphere than in the northern hemisphere. Such differences could translate into systematic variations in the predicted UCD surface density between fields at positive and negative Galactic latitudes. We do not attempt to model these potential asymmetries and instead retain fixed values of $R_{\odot}$ and $Z_{\odot}$ in our fiducial model. Consequently, uncertainties in the local Galactic structure, including possible north--south asymmetries, may contribute to the systematic uncertainty in the predicted UCD contamination.

An additional limitation is stellar multiplicity, as our model assumes that UCDs are single objects. In general, low-mass stars have lower companion fractions than solar-type stars \citep{Duchene2013}. Late-M and L dwarfs have binary fractions of $\sim10\%$ \citep{Factor2023}, while the fraction decreases to $\sim5\%$ for late-T and Y dwarfs \citep{Fontanive2018}. UCD binaries also tend to have near-equal mass components \citep{Burgasser2007,Fontanive2018}, and many have separations too small to be resolved in typical imaging \citep{Bouy2008,Laithwaite2020}. Unresolved binaries can therefore appear up to approximately 1 magnitude brighter than their individual components, potentially altering both the magnitude distribution and the overall number of UCDs entering a magnitude-limited sample. However, given the relatively low binary fraction of UCDs, we expect this effect to be modest.

Overall, these effects are not expected to dominate the uncertainty in the predicted UCD contamination. Stellar streams are unlikely to contribute substantially at the depths of the surveys considered here, while spiral arms, north--south asymmetries, and unresolved multiplicity are expected to introduce relatively modest changes to the predicted contamination. We therefore expect the uncertainties associated with these omitted effects to be smaller than those explored in Section \ref{sec:contamrate_and_modelparam}, where variations in the model parameters can change the predicted contamination by factors of several.

\bibliography{bib}{}
\bibliographystyle{aasjournalv7}

\end{document}